\documentclass[11pt]{article}
\usepackage{jheppub}
\usepackage{amsmath}
\usepackage{amssymb}
\usepackage{braket}
\usepackage{bbold}
\usepackage{pifont}

\newcommand{\ba}{\begin{eqnarray}}
\newcommand{\ea}{\end{eqnarray}}
\newcommand{\be}{\begin{equation}}
\newcommand{\ee}{\end{equation}}

\newcommand{\gmn}{g_{\mu\nu}}
\newcommand{\fmn}{f_{\mu\nu}}

\newcommand{\hmn}{h_{\mu\nu}}
\newcommand{\Vmn}{V_{\mu\nu}}
\newcommand{\Gmn}{\mathcal{G}_{\mu\nu}}

\newcommand{\md}{\mathrm{d}}

\newcommand{\beqn}{\begin{eqnarray}}
\newcommand{\eeqn}{\end{eqnarray}}
\newcommand{\td}{\mathrm{d}}

\newcommand{\p}{\partial}

\newcommand{\nn}{\nonumber}
\newcommand{\Tr}{\mathrm{Tr}}

\newcommand{\X}{\mathbb{X}}

\def\ph{\phantom}

\newcommand{\Jaa}{\nabla_{\rho}\nabla_{\sigma}\,h^{\rho\sigma}}
\newcommand{\Jbb}{\square\,h}
\newcommand{\Iaa}{S^{\rho\sigma}\,\nabla_{\rho}\nabla_{\lambda}\,h^{\lambda}_{\ \sigma}}
\newcommand{\Ibb}{S^{\rho\sigma}\,\nabla_{\rho}\nabla_{\sigma}\,h}
\newcommand{\Icc}{S^{\rho\sigma}\,\square\,h_{\rho\sigma}}
\newcommand{\Haa}{[S^2]^{\rho\sigma}\,\nabla_{\rho}\nabla_{\lambda}\,h^{\lambda}_{\ \sigma}}
\newcommand{\Hbb}{[S^2]^{\rho\sigma}\,\nabla_{\rho}\nabla_{\sigma}\,h}
\newcommand{\Hcc}{[S^2]^{\rho\sigma}\,\square\,h_{\rho\sigma}}
\newcommand{\Gaa}{S^{\rho\sigma}\,S^{\mu\nu}\,\nabla_{\rho}\nabla_{\mu}\,h_{\sigma\nu}}
\newcommand{\Gbb}{S^{\rho\sigma}\,S^{\mu\nu}\,\nabla_{\rho}\nabla_{\sigma}\,h_{\mu\nu}}
\newcommand{\Faa}{[S^3]^{\rho\sigma}\,\nabla_{\rho}\nabla_{\lambda}\,h^{\lambda}_{\ \sigma}}
\newcommand{\Fbb}{[S^3]^{\rho\sigma}\,\nabla_{\rho}\nabla_{\sigma}\,h}
\newcommand{\Fcc}{[S^3]^{\rho\sigma}\,\square\,h_{\rho\sigma}}
\newcommand{\Eaa}{[S^2]^{\rho\sigma}\,S^{\mu\nu}\,\nabla_{\rho}\nabla_{\mu}\,h_{\sigma\nu}}
\newcommand{\Ebb}{[S^2]^{\rho\sigma}\,S^{\mu\nu}\,\nabla_{\rho}\nabla_{\sigma}\,h_{\mu\nu}}
\newcommand{\Ecc}{[S^2]^{\rho\sigma}\,S^{\mu\nu}\,\nabla_{\mu}\nabla_{\nu}\,h_{\rho\sigma}}
\newcommand{\Daa}{[S^2]^{\rho\sigma}\,[S^2]^{\mu\nu}\,\nabla_{\rho}\nabla_{\mu}\,h_{\sigma\nu}}
\newcommand{\Dbb}{[S^2]^{\rho\sigma}\,[S^2]^{\mu\nu}\,\nabla_{\rho}\nabla_{\sigma}\,h_{\mu\nu}}
\newcommand{\Caa}{[S^3]^{\rho\sigma}\,S^{\mu\nu}\,\nabla_{\rho}\nabla_{\mu}\,h_{\sigma\nu}}
\newcommand{\Cbb}{[S^3]^{\rho\sigma}\,S^{\mu\nu}\,\nabla_{\rho}\nabla_{\sigma}\,h_{\mu\nu}}
\newcommand{\Ccc}{[S^3]^{\rho\sigma}\,S^{\mu\nu}\,\nabla_{\mu}\nabla_{\nu}\,h_{\rho\sigma}}
\newcommand{\Baa}{[S^3]^{\rho\sigma}\,[S^2]^{\mu\nu}\,\nabla_{\rho}\nabla_{\mu}\,h_{\sigma\nu}}
\newcommand{\Bbbb}{[S^3]^{\rho\sigma}\,[S^2]^{\mu\nu}\,\nabla_{\rho}\nabla_{\sigma}\,h_{\mu\nu}}
\newcommand{\Bcc}{[S^3]^{\rho\sigma}\,[S^2]^{\mu\nu}\,\nabla_{\mu}\nabla_{\nu}\,h_{\rho\sigma}}
\newcommand{\Aaa}{[S^3]^{\rho\sigma}\,[S^3]^{\mu\nu}\,\nabla_{\rho}\nabla_{\mu}\,h_{\sigma\nu}}
\newcommand{\Abb}{[S^3]^{\rho\sigma}\,[S^3]^{\mu\nu}\,\nabla_{\rho}\nabla_{\sigma}\,h_{\mu\nu}}

\newcommand{\Ja}{J_1}
\newcommand{\Jb}{J_2}
\newcommand{\Ia}{I_1}
\newcommand{\Ib}{I_2}
\newcommand{\Ic}{I_3}
\newcommand{\Ha}{H_1}
\newcommand{\Hb}{H_2}
\newcommand{\Hc}{H_3}
\newcommand{\Ga}{G_1}
\newcommand{\Gb}{G_2}
\newcommand{\Fa}{F_1}
\newcommand{\Fb}{F_2}
\newcommand{\Fc}{F_3}
\newcommand{\Ea}{E_1}
\newcommand{\Eb}{E_2}
\newcommand{\Ec}{E_3}
\newcommand{\Da}{D_1}
\newcommand{\Db}{D_2}
\newcommand{\Ca}{C_1}
\newcommand{\Cb}{C_2}
\newcommand{\Cc}{C_3}
\newcommand{\Ba}{B_1}
\newcommand{\Bb}{B_2}
\newcommand{\Bc}{B_3}
\newcommand{\Aa}{A_1}
\newcommand{\Ab}{A_2}

\newcommand{\gf}{{\mathfrak{F}}}
\newcommand{\scah}{\aleph}

\newcommand{\Jamn}{h^{\mu\nu}}
\newcommand{\Jbmn}{g^{\mu\nu}\,h}
\newcommand{\Iamn}{S^{\mu\sigma}\,h^{\nu}_{\ph\nu\sigma}}
\newcommand{\Ibmn}{S^{\mu\nu}\,h}
\newcommand{\Icmn}{g^{\mu\nu}\,S^{\rho\sigma}\,h_{\rho\sigma}}
\newcommand{\Hamn}{[S^2]^{\mu\sigma}\,h^{\nu}_{\ph\nu\sigma}}
\newcommand{\Hbmn}{[S^2]^{\mu\nu}\,h}
\newcommand{\Hcmn}{g^{\mu\nu}\,[S^2]^{\rho\sigma}\,h_{\rho\sigma}}
\newcommand{\Gamn}{S^{\mu\rho}\,S^{\nu\sigma}\,h_{\rho\sigma}}
\newcommand{\Gbmn}{S^{\mu\nu}\,S^{\rho\sigma}\,h_{\rho\sigma}}
\newcommand{\Famn}{[S^3]^{\mu\sigma}\,h^{\nu}_{\ph\nu\sigma}}
\newcommand{\Fbmn}{[S^3]^{\mu\nu}\,h}
\newcommand{\Fcmn}{g^{\mu\nu}\,[S^3]^{\rho\sigma}\,h_{\rho\sigma}}
\newcommand{\Eamn}{[S^2]^{\mu\rho}\,S^{\nu\sigma}\,h_{\rho\sigma}}
\newcommand{\Ebmn}{[S^2]^{\mu\nu}\,S^{\rho\sigma}\,h_{\rho\sigma}}
\newcommand{\Ecmn}{S^{\mu\nu}\,[S^2]^{\rho\sigma}\,h_{\rho\sigma}}
\newcommand{\Damn}{[S^2]^{\mu\rho}\,[S^2]^{\nu\sigma}\,h_{\rho\sigma}}
\newcommand{\Dbmn}{[S^2]^{\mu\nu}\,[S^2]^{\rho\sigma}\,h_{\rho\sigma}}
\newcommand{\Camn}{[S^3]^{\mu\rho}\,S^{\nu\sigma}\,h_{\rho\sigma}}
\newcommand{\Cbmn}{[S^3]^{\mu\nu}\,S^{\rho\sigma}\,h_{\rho\sigma}}
\newcommand{\Ccmn}{S^{\mu\nu}\,[S^3]^{\rho\sigma}\,h_{\rho\sigma}}
\newcommand{\Bamn}{[S^3]^{\mu\rho}\,[S^2]^{\nu\sigma}\,h_{\rho\sigma}}
\newcommand{\Bbmn}{[S^3]^{\mu\nu}\,[S^2]^{\rho\sigma}\,h_{\rho\sigma}}
\newcommand{\Bcmn}{[S^2]^{\mu\nu}\,[S^3]^{\rho\sigma}\,h_{\rho\sigma}}
\newcommand{\Aamn}{[S^3]^{\mu\rho}\,[S^3]^{\nu\sigma}\,h_{\rho\sigma}}
\newcommand{\Abmn}{[S^3]^{\mu\nu}\,[S^3]^{\rho\sigma}\,h_{\rho\sigma}}
\newcommand{\barM}{\bar{M}}
\newcommand{\barm}{\bar{m}}

\newcommand{\ccc}{u}
\newcommand{\ddd}{v}
\newcommand{\alphacoef}{\alpha}
\newcommand{\Lambdabak}{\Lambda}

\title{Massive graviton on arbitrary background: derivation, syzygies, applications}
\author[1]{Laura~Bernard,}
\author[1,2]{C\'edric~Deffayet,}
\author[1]{Mikael~von~Strauss}
\affiliation[1]{UPMC-CNRS, UMR7095,
 Institut d'Astrophysique de Paris, GReCO,\\
 98bis boulevard Arago, F-75014 Paris, France.}
\affiliation[2]{IHES, Institut des Hautes \'Etudes Scientifiques,\\
Le Bois-Marie, 35 route de Chartres, F-91440 Bures-sur-Yvette, France}
\emailAdd{bernard@iap.fr}
\emailAdd{deffayet@iap.fr}
\emailAdd{strauss@iap.fr}

\abstract{We give the detailed derivation of the fully covariant form of the quadratic action and the derived linear equations of motion for a massive graviton in an arbitrary background metric  (which were presented in arXiv:1410.8302 [hep-th]).
Our starting point is the de Rham-Gabadadze-Tolley (dRGT) family of ghost free massive gravities and using a simple model of this family, we are able to express this action and these equations of motion in terms of a single metric in which the graviton propagates, hence removing in particular the need for a "reference metric" which is present in the non perturbative formulation. We show further how 5 covariant constraints can be obtained including one which leads to the tracelessness of the graviton on flat space-time and removes the Boulware-Deser ghost. This last constraint involves powers and combinations of the curvature of the background metric. 
The 5 constraints are obtained for a background metric which is unconstrained, i.e. which does not have to obey the background field equations.
We then apply these results to the case of Einstein space-times, where we show that the 5 constraints become trivial, and Friedmann-Lema\^{\i}tre-Robertson-Walker space-times, for which we correct in particular some results that appeared elsewhere. 
To reach our results, we derive several non trivial identities, syzygies, involving the graviton fields, its derivatives and the background metric curvature. These identities have their own interest. We also discover that there exist backgrounds for which the dRGT equations cannot be unambiguously linearized.}


\begin{document} 
\maketitle
\flushbottom

\section{Introduction}
The last 15 years have seen several developments about massive gravity originating in the DGP model \cite{Dvali:2000hr} and its ability to produce a late time acceleration of the Universe via a large distance modification of gravity \cite{Deffayet:2000uy,Deffayet:2001pu}. One of the main motivations for massive gravity (see \cite{Rubakov:2008nh,Hinterbichler:2011tt,deRham:2014zqa} for recent reviews) is indeed the wish to replace dark energy by a non vanishing graviton mass, and to give a mass to the graviton is one of the simplest (conceptually speaking) ways to modify gravity in the Infra-Red. In fact, there has been many attempts to write a consistent theory for a massive graviton since the invention of general relativity and various problems have been encountered on this road. The unique consistent theory for a free (non self interacting) massive spin-2 field on a Minkowski space-time has been known for a long time as the Fierz-Pauli theory \cite{Fierz:1939ix}. It propagates 5 degrees of freedom of positive energy, those of a transverse, traceless, symmetric, two times covariant tensor. However, one of these d.o.f., a scalar mode, leads to the well known van-Dam Veltman Zakharov discontinuity: namely the fact that, however small the graviton mass, Fierz-Pauli theory leads to different physical predictions (such as light bending) from those of linearized General Relativity \cite{vanDam:1970vg}. This is enough to rule out Fierz-Pauli theory {\it stricto sensu} from solar system experiment. However, a way out was suggested by Vainshtein \cite{Vainshtein:1972sx} considering 
 self-interactions of the massive graviton. This so-called "Vainshtein mechanism" was criticized soon after it was proposed by Boulware and Deser in the seminal paper \cite{Boulware:1973my} and it was only recently that it was actually proven to operate as predicted \cite{Deffayet:2001uk,Babichev:2009us,Babichev:2009jt,Babichev:2010jd} (see also the review \cite{Babichev:2013usa}). Boulware and Deser also discovered a new pathology of generic massive gravities: 
the propagation of a ghost-like 6th degree of freedom at the nonlinear level. It was long thought impossible to obtain a massive gravity theory devoid of this ghost (see e.g. \cite{Boulware:1973my,Creminelli:2005qk}). However, a family of massive gravity theories was recently proposed by de Rham, Gabadadze, and Tolley (dRGT in the following) \cite{deRham:2010kj,deRham:2010ik,deRham:2011rn} in which the absence of ghost was first addressed in the so-called decoupling limit \cite{deRham:2010ik} (using, in particular, the approach of Refs. \cite{Creminelli:2005qk,ArkaniHamed:2002sp,Deffayet:2005ys}) and then fully shown at the nonlinear level\footnote{Note, however, that these results were initially debated \cite{Alberte:2010qb,Chamseddine:2011mu}.} by a Hamiltonian analysis, later extended to bimetric theories \cite{Hassan:2011ea,Hassan:2011zd,Hassan:2011tf,Hassan:2011hr} (see also \cite{Kluson,OthersCounting}).  The Hamiltonian analysis of these models remains however involved and does not clarify the reasons behind their soundness. 
The original discussions of the dRGT models have been using a metric formulation, however it has been pointed out that a vierbein formulation (first obtained in \cite{Hinterbichler:2012cn} building in particular on the older work of ref.~\cite{Nibbelink:2006sz}) makes easier the obtention of the constraint necessary to remove the Boulware Deser ghost \cite{Hinterbichler:2012cn,Deffayet:2012nr}\footnote{See also \cite{Hassan:2012wt,Mourad:2014xla,Mourad:2014roa,Mourad:2013rwa,Deser:2014hga}.} while it also makes the theory more elegant by the removal of an unpleasant matrix square root. Note however that the two formulations are not totally equivalent \cite{Deffayet:2012zc,Banados:2013fda} and we will consider here only the metric formulation.  

The aim of this paper is twofold. First, starting from the dRGT models, we provide the detailed derivation of the action and derived field equations of a massive graviton with (at most) 5 polarizations on an arbitrary background.
The obtained theory is written in a fully covariant way and has been presented in Ref.~\cite{Bernard:2014bfa}. In particular, we will, for a subset of dRGT models, show how the extraneous reference metric (usually taken to be Minkowski) can be fully integrated out, leaving only one metric which serves as a background for the graviton propagation. The obtained equations contain curvature tensors of the background metric entering into the mass term in a non trivial way. 
Secondly, we provide a full Lagrangian covariant proof that the graviton we consider has {\it always}, i.e. on arbitrary backgrounds, at most five polarizations\footnote{It is known that some polarizations can be prevented to propagate depending on the background \cite{Gumrukcuoglu:2011zh,DeFelice:2012mx}}
and not 6 as could be expected from the Boulware-Deser argument. To do so, we show how the field equations lead to 5 covariant constraints like in the long known case of Fierz-Pauli theory over flat space-time or Einstein space-times \cite{Espace,Buchbinder:1999ar,Higuchi:1986py,Porrati:2000cp,ArkaniHamed:2002sp,Deffayet:2011uk}. Out of these constraints, four are easy to find, just coming from the Bianchi identities associated with the diffeomorphism invariance of the kinetic terms. The fifth constraint is a scalar constraint necessary to eliminate the Boulware-Deser ghost and, in the Fierz-Pauli case, leads to the tracelessness of the graviton field. We will show that such a constraint can always be obtained irrespectively of the background metric, i.e. even when this background metric does not solve the background field equations. Such a constraint has been obtained in the generic case in the vierbein formulation of dRGT (with a Minkowski reference metric though) where it appears surprisingly simple, at least in the simplest models of the dRGT family. As we will see, the derivation of the similar constraint in the metric formalism and from linearized field equations is much more involved and requires several non trivial technical tricks. In particular, to reach our results, we derive several non trivial identities -~syzygies~- involving the graviton field, its derivatives and the background metric curvature. These identities have their own interest. We also discover that there exist backgrounds for which the equations cannot be unambiguously linearized.

This paper is organized as follows. In the following section we introduce massive gravity and covariant d.o.f. counting. Section \ref{sec:outline} summarizes the main results of the paper as well outlines, without giving the technical details, the general reasoning and technical steps. The detailed proof is given in the following two sections, while some long formulae and technical steps are differed to appendices. Section \ref{Applic} discusses some simple applications of the obtained results: flat and Einstein space-times as well as FLRW space-times. A last section contains conclusions.

\section{Introduction to massive gravity and dRGT theories}
\label{S1}
\subsection{Fierz-Pauli theory}
Let us first introduce the (quadratic) Fierz-Pauli theory \cite{Fierz:1939ix} on a flat background. This theory can be defined\footnote{For simplicity, we only discuss in this section the case with $d=4$ dimensions, but such a theory can be introduced in a similar way for arbitrary $d$, where it is found that a massive graviton in $d$ dimensions, as described by the quadratic Fierz-Pauli theory, has $(d-2)(d+1)/2$ physical polarizations, i.e. the same number of polarizations as a massless graviton in $d+1$ dimensions.} by the following action (in the absence of matter source) for a rank-2 covariant tensor $h_{\mu \nu}$ 
\ba \label{PF}
S_{h,m} & =& - \barM_h^2 \int d^4x \left[ \left(\partial_\mu h_{\nu\rho} \right)^2 - \left(\partial_\mu h\right)^2 \nonumber
+ 2 \left(\partial_\mu h\right)\left( \partial^\nu h^\mu_\nu \right)- 2 \left(\partial_\mu h_{\nu \rho}\right) \left(\partial^\nu h^{\mu \rho}\right)\right.\nonumber \\
&&\left. + \barm^2\left( h_{\mu \nu}h^{\mu \nu}-h^2\right)\right]\,.
\ea 
Here $\barM_h$ is a mass parameter, all indices are moved up and down with a flat canonical Minkowski metric $\eta_{\mu \nu}$, and $h$ is defined by $h \equiv h_{\mu \nu} \eta^{\mu \nu}$. The terms with derivatives on the right hand side of Eq. (\ref{PF}) are obtained by expanding the Einstein-Hilbert Lagrangian density $\sqrt{-g} R(g)$ at quadratic order around a flat metric, writing $g_{\mu \nu} = \eta_{\mu \nu} + h_{\mu \nu}$.
The mass terms appear with a factor of $\barm^2$ on the right hand side of Eq. (\ref{PF}), and this particular combination of $h^2$ and $(h_{\mu \nu})^2$ is the only one able to give a mass to the graviton $h_{\mu \nu}$ in a consistent and ghost-free way.  Note that this theory explicitly breaks general covariance and also that it uses two rank-2 covariant tensors, $h_{\mu \nu}$ as well as $\eta_{\mu \nu}$ which serves as a background on which $h_{\mu \nu}$ propagates. Varying the above action (\ref{PF}) with respect to $h_{\mu \nu}$ one easily obtains the field equations in vacuum
\ba \label{FIELDgen}
\delta \bar{E}_{\mu \nu} = 0\,,
\ea
where $\delta \bar{E}_{\mu \nu}$ is defined by 
\ba \label{FlatFE1}
\delta \bar{E}_{\mu \nu} & \equiv &  \delta \bar{\cal G}_{\mu \nu} + \frac{\barm^2}{2} \left(h_{\mu \nu} - h \; \eta _{\mu \nu}\right) \\ 
&\equiv & -\frac{1}{2}\left(\partial_\mu \partial_\nu h + \Box h_{\mu \nu} - \partial_\rho \partial_\mu h^\rho_\nu -\partial_\rho \partial_\nu h^\rho_\mu  + \eta_{\mu \nu}(\partial^\rho \partial^\sigma h_{\rho \sigma} - \Box h) \right) +  \frac{\barm^2}{2} \left(h_{\mu \nu} - h \; \eta _{\mu \nu}\right)\,.\nn
\ea
Here $ \delta \bar{\cal G}_{\mu \nu}$  is just the linearized Einstein tensor and hence, as a consequence of Bianchi identities, its divergence $\partial^\mu$ vanishes identically. Thus, acting with $\partial^\mu$ on Eq.  (\ref{FIELDgen}), we get (for a non vanishing $\barm$)
\ba \label{EQAUX1}
 \partial^\mu h_{\mu \nu} - \partial_\nu h = 0\,.
\ea
Taking another derivative of this equation yields
\ba \label{EQAUX2}
\partial^\nu \partial^\mu h_{\mu \nu} - \Box h = 0\,,
\ea 
where the combination appearing in the left hand side above happens to be just the linearization of the Ricci scalar.
This combination hence also appears taking the trace of the tensor $ \delta \bar{\cal G}_{\mu \nu}$ which appears in the field equations. This means that one has the following fundamental identity (where appropriate coefficients have been included)
\ba \label{consMin1}
2 \partial^\mu \partial^\nu \delta \bar{E}_{\mu \nu} + \barm^2 \eta^{\mu \nu} \delta \bar{E}_{\mu \nu} \sim 0\,,
\ea
where here and in the rest of this paper two expressions separated by the symbol  "$\sim$" are by definition equal {\it off-shell} (i.e. without using the field equation) up to (possibly vanishing) terms containing undifferentiated or once differentiated $h_{\mu \nu}$ but no second or higher derivatives acting on $h_{\mu \nu}$. 
Explicitly one has off-shell 
\ba \label{consMin2}
2 \partial^\mu \partial^\nu \delta \bar{E}_{\mu \nu} + \barm^2 \eta^{\mu \nu} \delta \bar{E}_{\mu \nu} = -\frac{3}{2} \barm^4 h\,.
\ea
Now the left hand side of Eq.(\ref{consMin1}) (or (\ref{consMin2})) vanish on-shell which in turn yields a constraint on $h_{\mu \nu}$, specifically that $h$ vanishes in vacuum. 
Thus, using also (\ref{EQAUX1}), we conclude that $h_{\mu \nu}$ is transverse and traceless in vacuum, i.e. that it obeys the two equations 
\ba
\partial^\mu h_{\mu \nu} &=&  0\,, \label{CONS1}\\
h&=& 0   \label{CONS2}\,.
\ea
These two equations  give $5$ Lagrangian constraints (being at most first order in derivatives)  and this removes $5$ of the {\it a priori} $10$ dynamical degrees of freedom of $h_{\mu \nu}$. Hence, 
quadratic Fierz-Pauli theory propagates 5 polarizations. A similar conclusion can also be reached using a Hamiltonian counting (see e.g. \cite{Boulware:1973my}). 
Here, we stress that the crucial step to reach this conclusion is the identity (\ref{consMin1}) which would not hold for any other mass term than the one of Fierz-Pauli. On the other hand, the four constraints 
(\ref{EQAUX1}) are consequences of the Bianchi identities leading to 
\ba
\partial^\mu \delta \bar{E}_{\mu \nu} \sim 0\,.
\ea
And hence, constraints analogous to (\ref{EQAUX1}) can always be found.
Note further, that Fierz-Pauli theory and the above described way of counting its degrees of freedom can easily be extended to the case of a background metric describing a maximally symmetric space-time with a non vanishing curvature \cite{Porrati:2000cp} or even a generic Einstein space-time \cite{Deffayet:2011uk}.

\subsection{Nonlinear massive gravity}
In order to go beyond Fierz-Pauli, this needs to be suitably nonlinearly extended.  
Such a nonlinear extension can generically be formulated with an action of the form 
\ba \label{NLPFG} 
S_{g,m} = M_g^2 \int d^4 x \sqrt{|g|} \left[R(g) - 2 m^2 V\left({\gf}\right)\right]\,,
\ea
where $V$ is a suitably chosen scalar function of ${\gf}^\mu_{\hphantom{\mu} \nu} \equiv g^{\mu \sigma} f_{\sigma \nu}$, $m$ and $M_g$ are mass parameters, $R(g)$ is the Ricci scalar constructed from the metric $g_{\mu \nu}$, and the theory contains, besides the dynamical metric $g_{\mu \nu}$, a non dynamical metric $f_{\mu \nu}$ usually considered to be flat, and which appears in the action only through the tensor $\gf$. Note that since in the considered theory space-time is endowed with two metrics, there is some ambiguities on how to move indices. In the rest of this paper, all indices will always be moved up and down with the dynamical metric $\gmn$. If one wants to consider a "nonlinear massive gravity" {\it stricto sensu} (i.e. an nonlinear extension of Fierz-Pauli theory) the potential $V$ should be chosen such that (i) $g_{\mu \nu} = \eta_{\mu \nu}$ is a solution of the field equations (irrespectively of the choice of $f_{\mu \nu}$), and (ii) when expanded at quadratic order around this flat background, the action (\ref{NLPFG}) has the Fierz-Pauli form (\ref{PF})\footnote{Note that when one has two metrics, one can write any non trivial non derivative invariant built from the metrics as a function of $\gf$, and hence the only restriction (besides diffeomorphism invariance) comes here from requirements (i) and (ii) on $V$.}. It is easy to figure out that there are infinitely many functions $V$ that satisfy these requirements (see e.g. \cite{Damour:2002ws}). The dRGT theory has an action of the above form (\ref{NLPFG}) depending (in 4 dimensions) on 4 reals parameters $\beta_n$, a subset of which can be chosen in order to fulfill the conditions (i) and (ii) above. However, we stress that our results will be true whether or not this fulfilment is chosen, and hence do not only apply to nonlinear massive gravity {\it stricto sensu} as it is defined above, but to larger set of theories.

As first noticed by Boulware and Deser \cite{Boulware:1973my}, quadratic Fierz-Pauli theory (\ref{PF}) and its nonlinear version (\ref{NLPFG}) differ in general dramatically as far as the number of propagating degrees of freedom is concerned. Indeed, a generic nonlinear massive gravity as defined in (\ref{NLPFG}) propagates an additional sixth ghost-like polarization, usually called the Boulware-Deser ghost. 
Schematically, this comes from the fact that the analog of the constraint (\ref{CONS2}) is lost, while there are still $4$ constraints similar to (\ref{CONS1}) (or (\ref{EQAUX1})) (see e.g. 
\cite{Deffayet:2005ys}).   Indeed, now vary action (\ref{NLPFG}) with respect to $g_{\mu \nu}$ to obtain the field equations,
\be\label{g_eom1_first}
E_{\mu\nu} \equiv \mathcal{G}_{\mu\nu}+m^2\,V_{\mu\nu}=0\,,
\ee
where ${\cal G}_{\mu \nu}$, the Einstein tensor built from the metric $\gmn$, and the interaction contribution $V_{\mu\nu}$ are both defined through
\be\label{GVdefs_first}
\mathcal{G}_{\mu\nu}=R_{\mu\nu}-\frac{1}{2}\gmn R\,,\qquad
V_{\mu\nu}\equiv\frac{-2}{\sqrt{|g|}}\frac{\delta(\sqrt{|g|}V)}{\delta g^{\mu\nu}}\,.
\ee
$V_{\mu \nu}$ is obtained from varying the term coupling the two metrics $\sqrt{-g} V({\gf})$, and hence contains no derivatives. 
As a consequence of that and of the usual Bianchi identities we have that 
\ba 
\nabla^{\mu} E_{\mu \nu} \sim 0\,, 
\ea
where $\nabla^\mu$ is the covariant derivative built from the metric $g_{\mu\nu}$. The above equation, yields, taking into account the field equations, the four Lagrangian constraints (analogous to (\ref{CONS1}))
\ba \label{CONSNL}
\nabla^{\mu} E_{\mu \nu} =m^2 \nabla^{\mu} V_{\mu \nu}= 0\,. 
\ea
However, on tracing over Eq.(\ref{g_eom1_first}) and using derivatives of Eq.(\ref{CONSNL}), there is now (as opposed to linear Fierz-Pauli theory) no reason in general to get an extra constraint (cf. also \cite{Deffayet:2005ys}). This reflects the fact that it was thought impossible to construct a nonlinear Fierz-Pauli theory, with a suitable potential $V$, devoid of the Boulware Deser mode \cite{Boulware:1973my,Creminelli:2005qk} until the discovery of de Rham-Gabadadze-Tolley (henceforth dRGT) theories that we now introduce, introducing also some notations used in later sections.

\subsection{dRGT theories}
dRGT theories are nonlinear Fierz-Pauli theories (\ref{NLPFG}) for which the function $V$ takes a special form. Using the parametrization of dRGT theories proposed in Refs. \cite{Hassan:2011hr,Hassan:2011vm}, the action 
is containing a square root $S$ of the tensor $\gf$ defined as\footnote{This is a matrix square root. In fact a tensor like $\gf$ can have finitely many, infinitely many, or no square roots. See e.g. \cite{Deffayet:2012zc}.} 
\ba \label{Sdef} 
S^\mu_{\hphantom{\mu} \sigma}S^\sigma_{\hphantom{\sigma} \nu} = g^{\mu \sigma} f_{\sigma \nu} = \gf^{\mu}_{\hphantom{\mu} \nu}\,,
\ea
and can be written,
\be\label{SdRGT} 
S_{g,m} = M_g^2 \int\td^4 x \sqrt{|g|} \left[R(g) -  2m^2 V\left(S;\beta_n\right)\right]\,,
\ee
where the potential reads 
\be\label{Vdef}
V(S;\beta_n)=\sum_{n=0}^3\beta_ne_n(S)\,,
\ee
where the $\beta_n$ are 4 dimensionless parameters and $e_n(S)$ are $n$'th order elementary symmetric polynomials of the eigenvalues of their matrix argument. For a generic $d\times d$ matrix $\mathbb{M}$, $e_n(\mathbb{M})$  is given by,
\be
e_n(\mathbb{M}) = \frac{1}{n!} \mathbb{M}^{a_1}{}_{[a_1}...\mathbb{M}^{a_n}{}_{a_n]}\,, 
\ee
where the brackets $[\;]$ over indices denote the sum over unnormalized antisymmetrized permutations. Note that in $d$ dimensions, $e_n(S)=0$ for any $n>d$ while $e_d(S)=\det(S)$. They can also be constructed iteratively starting from $e_0(S)=1$ and using the recursive definition,
\be\label{edef}
e_n(S)=-\frac{1}{n}\sum_{k=1}^{n}(-1)^k\Tr[S^k]\,e_{n-k}(S)\,,\qquad n\geq1\,,
\ee
with $\Tr[X]=X^\rho_{~\rho}$ indicating a matrix trace operation, and $S^k$ is the k-th power of the tensor $S$ (considered as a matrix), i.e. 
\be \label{powerdef}
\left[ S^k \right]^\mu_{\hphantom{\mu} \nu} = S^\mu_{\hphantom{\mu} {\rho_1} } S^{\rho_1}_ {\hphantom{\rho_1} \rho_2 } \cdots S^{\rho_{k-1}}_{\hphantom{\rho_{k-1}} \nu }\,.
\ee
We stress here that above and in the rest of the paper, we consider at various places matrix operations (such as the elevation to a power as above) as applied to a tensor such as $S$, which, when we do so, must be taken to have one contravariant space-time index (which then corresponds to the line index of the matrix) and one covariant index (corresponding to the column of the matrix).

An alternative formulation to action \eqref{SdRGT} is to consider $S$ as an independent field and to enforce the relation \eqref{Sdef} by a Lagrange multiplier $C_{\mu}^{\ph\mu\nu}$, adding to the Lagrangian a term of  the form (in the spirit of e.g.~ Ref.~\cite{deRham:2010kj}),
\be\label{conslag}
 \sqrt{|g|}\,C_{\mu}^{\ph\mu\nu}\left(S^\mu_{\hphantom{\mu} \sigma}S^\sigma_{\hphantom{\sigma} \nu} - g^{\mu \sigma} f_{\sigma \nu}\right)\,.
\ee
This alternative does not feature the presence of the unpleasant square root in the action. Whichever way is chosen, the presence of the square root either directly in the action or via a Lagrange multiplier is a somewhat inelegant aspect of the theories considered which also introduces technical and conceptual difficulties. A vielbein formulation of these theories \cite{Hinterbichler:2012cn} (or at least of a subset of them) offers a nice alternative which does not suffer from the same lack of elegance (even though the two formulations do not lead to equivalent theories \cite{Deffayet:2012zc,Banados:2013fda}). For the purposes of this paper we will however work in the metric formulation, using the action \eqref{SdRGT} and assume the existence of a tensor $S$ obeying \eqref{Sdef}.

As a step to derive the field equations, but also for a later use, we can express the variation $\delta e_n(S)$ of the $e_n(S)$  in terms of that of $S^2 \equiv \gf$ noted $\delta S^2 \equiv \delta \gf$
\be\label{d_e_n}
\delta e_n(S)=-\frac1{2}\sum_{k=1}^n(-1)^k\Tr[S^{k-2}\delta S^2]\,e_{n-k}(S)\,,\qquad n\geq1\,,
\ee
while obviously $\delta e_0(S)=0$ (since $e_0(S)=1$). 
This expression is obtained varying the recursive relation \eqref{edef} w.r.t.~$S$ and a detailed derivation of it is given in appendix \ref{deltaeninduc}. 
It holds provided that $S$ is invertible, which is always true whenever $f$ is ($g$ having obviously an inverse). The interest of this expression is that $\delta S^2$ can easily be obtained in terms of the variation of the metric $g_{\mu \nu}$ (or its inverse) - see the definition \eqref{Sdef}. Using (\ref{d_e_n}), it is easy to  work out the detailed form of the interaction contribution,
\be\label{Vmn1}
V_{\mu\nu}=g_{\mu\rho}\sum_{n=0}^3\sum_{k=0}^n(-1)^{n+k}\beta_n[S^{n-k}]^\rho_{\ph\rho\nu}\,e_k(S)\,,
\ee
which enters into the field equations (\ref{g_eom1_first}). In the original construction of \cite{deRham:2010kj,deRham:2010ik,deRham:2011rn} the reference metric was taken to be a flat Minkowski metric, i.e.~$\fmn=\eta_{\mu\nu}$. Out of the 4 $\beta_n$ parameters, one parameter was also fixed such that $\gmn=\eta_{\mu\nu}$ was a valid solution to the equations of motion and another was chosen such that $m^2$ corresponded to the Fierz-Pauli mass for the fluctuations $h_{\mu\nu}=\gmn-\eta_{\mu\nu}$ (condition (i) and (ii) given before). The choice of trading one $\beta_n$ parameter for the Fierz-Pauli mass can be made without loss of generality because it simply amounts to an overall scaling of the parameters, but the other fixing is more restrictive on the theory. However, it was shown in \cite{Hassan:2011tf}, using an ADM parametrization, that the choice of $\fmn=\eta_{\mu\nu}$ could be relaxed and that the theory is free of the Boulware-Deser mode for arbitrary $\fmn$.\footnote{The dRGT theories seem to suffer from other problems though, ranging from difficulties in agreeing with cosmological observations to generic instabilities, hidden strong coupling and causality issues \cite{Gumrukcuoglu:2011zh,DeFelice:2012mx,Gratia:2013uza,Comelli:2014bqa,Burrage:2011cr, Burrage:2012ja,Gruzinov:2011sq,Deser:2013qza,Deser:2013eua,Deser:2012qx,Deser:2014hga,Yu:2013owa}.
} This allows to get flat space-time solutions for $\gmn$ without having to fix any $\beta_n$ as can be seen by considering solutions of the form $\fmn=c^2\gmn$. In this case the interaction term \eqref{Vmn1} reduces to a cosmological constant,
\be
V_{\mu\nu} = \Lambda\,\gmn\,,\qquad\Lambda=\beta_0+3c\beta_1+3c^2\beta_2+c^3\beta_3\,.
\ee
Hence, it is generically possible to set $\Lambda=0$ and thus have flat $\gmn$ solutions without fixing a parameter of the theory but instead allowing for $c\neq1$ \cite{Hassan:2012wr}. This leaves three free interaction parameters as opposed to the two appearing in the original dRGT construction. Note however that, as can easily be seen from \eqref{Vdef} or \eqref{Vmn1}, $\beta_0$ simply parametrizes a bare cosmological constant, which by itself does not give any mass to the graviton. 

\subsection{The "$\beta_1$ model" of dRGT family}
\label{sec:b1}
In the following, we will consider the simple case where $\beta_2$ and $\beta_3$ vanish but will keep otherwise $\beta_0$ and $\beta_1$ arbitrary as well as the metric $f_{\mu \nu}$ (which will in particular not assumed to be covering a Minkowski space-time). We will call the corresponding model, the "$\beta_1$ model", since for it to be non trivial, $\beta_1$ needs to be non vanishing which we will also assume (but no further assumption will be made on $\beta_0$), otherwise one is left with a mere cosmological constant. As we just discussed, a generic $\beta_1$ model will not necessarily fulfill conditions (i) and (ii) above (e.g. in the case where $\beta_0$ vanishes in addition to 
$\beta_2$ and $\beta_3$, there is no solution of the field equations where $g_{\mu \nu}$ is flat irrespectively of the choice of $f_{\mu \nu}$), our results will hence be valid for theories not in the category of "nonlinear massive gravity" {\it stricto sensu}. For the $\beta_1$ model, the field equations are given by 
\be \label{beta1field}
\mathcal{G}_{\mu\nu}+m^2\left[\beta_0\,\gmn+\beta_1\,g_{\mu\rho}\left(
e_1(S)\delta^\rho_{\nu}-S^\rho_{~\nu}\right)\right]=0\,,
\ee
where we recall that $e_1(S)$ is just the trace of $S$, i.e.~$e_1(S)= S^\rho_{~\rho}$. We observe that these equations are linear in $S$, so by tracing the equations we find that,
\be
-R+m^2\left[4\beta_0+3\beta_1\,e_1(S)\right]=0\,,
\ee
which we may use in turn to get the following expression of $S$ in terms of Ricci curvatures of $\gmn$.
\be\label{b1_Ssol}
S^\rho_{~\nu}=\frac1{\beta_1m^2}\left[R^\rho_{~\nu}-\frac1{6}\delta^\rho_{\nu} R
-\frac{m^2\beta_0}{3}\,\delta^\rho_{\nu}\right]\,.
\ee
 Since $S^\rho_{~\sigma}S^\sigma_{~\nu} = g^{\rho\sigma}f_{\sigma\nu}$, this can equivalently be seen (by squaring the above equation) as expressing $\fmn$ in terms of $\gmn$ and its Ricci curvatures (see eq.~(\ref{fmnRicci})). This remarkable feature means that later on, in our study of perturbation theory, we will be able to eliminate any and all occurences of the auxilliary metric $\fmn$ in favor of $\gmn$ and its curvature using the above equations. The resulting perturbative equations will then describe the dynamics of a spin-2 field propagating on a generic background $\gmn$ without any need for a second reference metric. This feature is unique to the $\beta_1$ model and only requires a non-vanishing $\beta_1$.

\section{Outline of the general reasoning and results}\label{sec:outline}

Before going through the technical details of what is a rather cumbersome analysis we will first sketch the general procedure and technical steps of our reasoning, stressing in particular where the technical difficulties arise. This will hopefully help in making the subsequent analysis more transparent and easier to follow, as well as allow the reader to have a first overview of our achievements without having to enter into the details. For the sake of simplicity, we discuss in this section only the $\beta_1$ model, however some of our results presented later are more general and apply to an arbitrary model in the dRGT family. 

We start from the field equations of the $\beta_1$ model (i.e. Eq.~\eqref{g_eom1_first}, with $V_{\mu \nu}$ given in (\ref{Vmn1}) with $\beta_2$ and $\beta_3$ vanishing),
\be\label{g_eom1p}
E_{\mu\nu}=\mathcal{G}_{\mu\nu}+m^2\,V_{\mu\nu}=\mathcal{G}_{\mu\nu}+m^2\left[\beta_0\,\gmn+\beta_1\,g_{\mu\rho}\left(
e_1(S)\delta^\rho_{\nu}-S^\rho_{~\nu}\right)\right]=0\,.
\ee
Then, we linearize these equations around some solution for $g_{\mu \nu}$, defining $h_{\mu \nu}$ as the first order perturbation of the metric $g$ (the metric $\fmn$ being considered fixed, as usual). The obtained equations read  (with obvious notations)
\ba\label{dg_eom1p}
\delta E_{\mu\nu} &\equiv&   \delta\mathcal{G}_{\mu\nu}
+m^2\delta V_{\mu\nu}
\label{dEoM}
\equiv\left[\tilde{\mathcal{E}}_{\mu\nu}^{\ph\mu\ph\nu\rho\sigma}
+m^2\,\mathcal{M}_{\mu\nu}^{\ph\mu\ph\nu\rho\sigma}\right]h_{\rho\sigma}=0\,,
\ea
where $\tilde{\mathcal{E}}_{\mu\nu}^{\ph\mu\ph\nu\rho\sigma} h_{\rho\sigma}$ gives by definition the linearized Einstein tensor $\delta\mathcal{G}_{\mu\nu}$ and 
$\mathcal{M}_{\mu\nu}^{\ph\mu\ph\nu\rho\sigma} h_{\rho\sigma}$ the linearized interaction term $\delta V_{\mu\nu}$.
Obviously, $\delta\mathcal{G}_{\mu\nu}$  is easy to obtain. The difficult step consists in getting the "mass term" $\mathcal{M}_{\mu\nu}^{\ph\mu\ph\nu\rho\sigma} h_{\rho\sigma}$ and our first achievement is to provide here an explicit and covariant form for this term. 

In order to do so, a first necessary step consists in expressing the variation of $S$, $\delta S$, in terms of $h_{\mu \nu}$, since this term appears in the linearization of the field equation (\ref{g_eom1p}) above . The corresponding expression is given in Eq.~(\ref{d_S}). To do so, one can notice that $\delta S$ obeys the equation (see (\ref{Sdef}))\footnote{This equation has also been considered in Ref.\cite{Kodama:2013rea} for the study of perturbations over the Schwarzschild-de Sitter black hole.}
\be \label{SylvdS}
S^\mu_{\hphantom{\mu} \nu} \left(\delta S\right)^\nu_{\hphantom{\nu} \sigma}+ \left(\delta S\right)^\mu_{\hphantom{\mu} \nu}S^\nu_{\hphantom{\nu} \sigma}=\delta \gf^\mu_{\hphantom{\mu} \sigma} \,.
\ee
which can be considered as a matrix equation in the class of the so-called Sylvester equation (see appendix \ref{app_sylv}), and where the right hand side is easy to get in terms of $h_{\mu \nu}$. A known mathematical results is that this equation has a unique solution for  $\delta S$ if and only if the spectrum of $S^\mu_{\hphantom{\mu} \nu}$ and $-S^\mu_{\hphantom{\mu} \nu}$ do no intersect (which is in general the case here). In this case, one can express the solution for $\delta S$ linearly in terms of $\delta \gf$, see e.g. \cite{Sylvesterpoly}. We will here rederive this result and in fact provide an alternative expression to the one we know from the mathematical literature \cite{Sylvesterpoly}. 

Having obtained the linearized field equations (\ref{dEoM}), it is possible, as we saw in the previous section, to replace there any occurence of $f_{\mu \nu}$ (or equivalently any occurence of $S$, since in fact $f$ 
only enter the field equation in the term $\mathcal{M}_{\mu\nu}^{\ph\mu\ph\nu\rho\sigma} h_{\rho\sigma}$ through $S$) by an expression containing only $\gmn$ and its curvature. The possibility to do this replacement singles out the $\beta_1$ model as rather special and is, apart from the mere simplicity of the model, a reason for why we have focussed on this model in the present work. Doing so we obtain the field equations for a graviton $h_{\mu \nu}$ which propagates in a single metric $g_{\mu \nu}$ entering into its kinetic term but also in its mass term where the Ricci curvatures of $g$ explicitly appear.  These equations can be considered as the starting point of the second part of our work which consists in proving covariantly that these equations propagates only at most 5 polarizations of $h_{\mu \nu}$ irrespectively of the metric $g$. In particular, we stress that in this proof, summarized in the rest of this section, we will not use the fact that $g$ is a solution of the background equations (\ref{g_eom1p}). 

As outlined in section \ref{S1} it is easy to find four vector constraints stemming from the fact that $\nabla^\mu {\delta {\cal G}_{\mu \nu}} \sim 0$ as a consequence of the Bianchi identity (where $\nabla$ denotes the covariant derivative taken with respect to the background metric $\gmn$ and the exact expression for $\nabla^\mu {\delta {\cal G}_{\mu \nu}}$ is given in Eq. (\ref{Bianpert})). 
In analogy with the flat space case considered in section \ref{S1} we are interested in finding a fifth scalar constraint which generalizes the constraint $h=0$ for the flat space case (c.f. Eq.~\eqref{CONS2}), as opposed to the vector constraints 
\ba \label{vectorconsgen}
\nabla^\mu\delta E_{\mu\nu}=0\,,
\ea 
which in turn generalize Eq.~\eqref{CONS1}. Accordingly, we look for a linear combination of scalars made by tracing over the field equations \eqref{dEoM} and their second derivatives which would not contain any derivatives of $h_{\mu \nu}$ of order strictly higher than one. However, we have now at hand two (symmetric) tensors which can be use to take traces, namely the metric $\gmn$ and its Ricci curvature $R_{\mu \nu}$. Equivalenty we could also use the metric and the tensor $S_{\mu \nu}$, trading $R_{\mu \nu}$ for $S_{\mu \nu}$ via equation (\ref{b1_Ssol}). We will in fact choose the second possibility which turns out to be more convenient technically speaking. We stress however that the two possibilities are strictly equivalent and do not impose any restriction on $g$, since (\ref{b1_Ssol}) can just be considered as a definition of $S_{\mu \nu}$ in term of $R_{\mu \nu}$ (as opposed to a  background field equation).  
Hence we define the  scalars $\Phi_i$ arising out of tracing the equations of motion
\be \label{defphi}
\Phi_i\equiv[S^i]^{\mu\nu}\,\delta E_{\mu\nu} \,,
\ee
together with the scalars $\Psi_i$ arising out of tracing the derivative of the divergence of the equations of motion in various ways,\footnote{The $i=0$ case includes the ordinary trace obtained by contracting the free indices together using $\gmn$ as a raising operator.}
\be \label{delpsi}
\Psi_{i}\equiv\frac{1}{2}\,[S^i]^{\mu \nu} \nabla_{\nu} \nabla^{\lambda}\,\delta E_{\lambda\mu}\,.
\ee
where the tensor $S^i$ is defined as in Eq. (\ref{powerdef}). 
These scalars actually exhaust all possibilities.
In fact this set of scalars can be further restricted by noticing that any power of $S$, $S^i$, with $i \geq 4$ can be replaced by a linear combination of lower powers of $S$ via the use of the Cayley-Hamilton theorem, as will be explained with more details below. To summarize we look for  a specific linear combination,
\be\label{cond_1}
\sum_{i=0}^{3}\left(\ccc_{i}\,\Phi_{i} + \ddd_{i}\,\Psi_{i}\right)\,,
\ee
with a set of seven\footnote{There are eight scalars $\{\ccc_i,\ddd_i\}$, but they need only to be determined up to an overall factor.} scalars $\{\ccc_i,\ddd_i\}$ to be determined, such that 
\be \label{lincombwesearch}
\sum_{i=0}^{3}\left(\ccc_{i}\,\Phi_{i} + \ddd_{i}\,\Psi_{i}\right) \sim 0\,,
\ee
i.e.~which contains off-shell no second (or higher) derivatives of $h_{\mu \nu}$. In the following, we will hence concentrate on the second derivatives of $h_{\mu \nu}$ appearing in the scalars $\Phi_{i}$ and $\Psi_{i}$ and show that these are in the form of linear combinations (with S-dependent coefficients) of 26 different scalars $\scah_i$ made by contracting $\nabla_\mu \nabla_\nu h_{\rho \sigma}$ with powers of $S$ (including zeroth power which just gives the metric). Examples of such scalars are $\Jaa$ and $\Abb$, and the exhaustive list \footnote{\label{footnotelist} All scalars made by contracting $\nabla_\mu \nabla_\nu h_{\rho \sigma}$ with powers of $S$ lower or equal to $3$ either appear explicitly in the list or are $\sim$ to scalar of this list as can be seen by commuting covariant derivatives acting on $\hmn$ - which only creates terms $\sim 0$ containing Riemann tensors of the background. Scalars made by contracting $\nabla_\mu \nabla_\nu h_{\rho \sigma}$ with powers of $S$ larger or equal to $4$ are linear combinations of those of the list as can be seen via the use of the Cayley-Hamilton theorem to replace such powers of $S$ by linear combinations of cubic and lower powers of $S$.} of them is given in Equations~\eqref{scalars1} to \eqref{scalars2}. Hence the linear combination (\ref{cond_1}) verifies
\be \label{linalphacoef}
\sum_{i=0}^{3}\left(\ccc_{i}\,\Phi_{i} + \ddd_{i}\,\Psi_{i}\right) \sim \sum_{i=0}^{26} \alphacoef_i \scah_i \,,
\ee
where the $\alpha_i$ are scalars functions of $\ccc_i, \ddd_i$, $S_{\mu \nu}$ and $\gmn$ that we will determine. Writing the equation (\ref{lincombwesearch}) we get {\it a priori} 26 equations for the seven unknowns $\{\ccc_i,\ddd_i\}$ by equating to zero each coefficient $\alphacoef_i$ of the  $\scah_i$. We will however show that not all the scalars  $\scah_i$ are independent of each other, thanks to identities, {\it syzygies}, which can be obtained using again the Cayley Hamilton theorem applied to a suitable matrix, as we will explain (this is a consequence of the "second fundamental theorem" of invariant theory \cite{Procesi,Sneddon,Hilbert}). These identities are gathered in appendix \ref{Syzygies} and they are just enough 
 to  reduce to seven the number of equations to be solved to fulfill (\ref{lincombwesearch}). This yields a unique solution for the coefficients $\{\ccc_i,\ddd_i\}$ which translates into the (we stress: off-shell) identity 
\begin{equation} \label{finalidentity}
\dfrac{m^2\,\beta_1\,e_4}{4}\,\Phi_0 \sim - e_3\,\Psi_0 + e_2\,\Psi_1-e_1\,\Psi_2 + \Psi_3  \,,
\end{equation}
where here and in the rest of this paper, we do not write out the functional dependence of the $e_n$ when they depend only on $S$, i.e.~it is to be understood that $e_n=e_n(S)$ (otherwise we will make the functional dependence explicit). Hence, using the field equations, we get the scalar constraint
\begin{equation} \label{scafin}
-\dfrac{m^2\,\beta_1\,e_4}{4}\,\Phi_0 - e_3\,\Psi_0 + e_2\,\Psi_1-e_1\,\Psi_2 + \Psi_3 = 0 \,.
\end{equation}
It can be seen that, when one sets $g_{\mu \nu}$ to be equal to the flat metric $\eta_{\mu \nu}$, this constraint is just $h=0$.  This not only shows that one recovers the constraint (\ref{CONS2}), as one should, but also that the constraint (\ref{scafin}) 
is independent of the vector constraints as it should. The flat space-time limit as well as the applications of these results to the case of Einstein and FLRW space-times are discussed in details in section \ref{Applic}.

\section{Linearized field equations for generic dRGT models}

In this section we obtain the linearized field equations for a generic model in the dRGT class \eqref{SdRGT} or equivalently the second variation of the action (in particular, we will not assume here the vanishing of the parameters $\beta_2$ and $\beta_3$). As we recalled above, this requires in particular computing the generic variation of a square root $\sqrt{\gf} \;(\equiv S)$ once the variation of $\gf \; (\equiv S^2)$ itself is known.

\subsection{Varying the equations of motion}
We aim here at giving an explicit covariant expression for the linearized field equations of dRGT theories which read  
\be\label{d_eom1}
\delta E_{\mu\nu} = \delta\Gmn+m^2\,\delta\Vmn=0\,.
\ee
where all quantities are computed at first order in the perturbation $\hmn$ of the dynamical metric $\gmn$ (henceforth providing a background over which the $\hmn$ propagate, and it is to be understood that all quantities that are not part of the perturbations are evaluated w.r.t. to this background solution $\gmn$), and the metric $f_{\mu \nu}$ is considered fixed (as it appears so in the action). 
Here, the variation of the Einstein tensor is given by,
\begin{align}
\label{dGmn}
\delta\Gmn
&=\mathcal{E}_{\mu\nu}^{\ph\mu\ph\nu\rho\sigma}h_{\rho\sigma}+\tfrac1{2}\left[
g_{\mu\nu} R^{\rho\sigma}-\delta^\rho_\mu \delta^\sigma_\nu R\right]h_{\rho\sigma}
\equiv \tilde{\mathcal{E}}_{\mu\nu}^{\ph\mu\ph\nu\rho\sigma}h_{\rho\sigma}\,,
\end{align}
where, for later considerations, we have defined
\begin{align}\label{EinstOp}
\mathcal{E}_{\mu\nu}^{\ph\mu\ph\nu\rho\sigma}h_{\rho\sigma}
\equiv -\tfrac1{2}\left[\delta^\rho_\mu\delta^\sigma_\nu\nabla^2+g^{\rho\sigma}\nabla_\mu\nabla_\nu 
-\delta^\rho_\mu\nabla^\sigma\nabla_\nu-\delta^\rho_\nu\nabla^\sigma\nabla_\mu
-g_{\mu\nu} g^{\rho\sigma}\nabla^2 +g_{\mu\nu}\nabla^\rho\nabla^\sigma\right]h_{\rho\sigma}\,,
\end{align}
which contains all terms with two derivatives acting directly on $\hmn$. We also implicitly defined $\tilde{\mathcal{E}}_{\mu\nu}^{\ph\mu\ph\nu\rho\sigma}$ which we used in our discussions of the previous section. The variation of the interaction contribution \eqref{Vmn1} can be obtained using Eq.~\eqref{d_e_n} and is given by,
\begin{align}\label{dVmn}
\delta V_{\mu\nu}
=&\,\frac1{2}\,h_{\mu\sigma}V^{\sigma}_{\ph\sigma\nu}
+\frac1{2}\, g_{\mu\rho}\delta[S^2]^\rho_{\ph\rho\sigma}\left[\beta_2\mathbb{1}
+\beta_3(e_1\mathbb{1}-S)\right]^\sigma_{\ph\delta\nu}\nn\\
&-\frac1{4}\, g_{\mu\lambda}\sum_{n=1}^3\sum_{k=1}^n\sum_{m=1}^k(-1)^{n+k+m}\beta_ne_{k-m}
[S^{n-k}]^\lambda_{\ph\lambda\nu}[S^{m-2}]^\sigma_{\ph\delta\rho}\delta[S^2]^\rho_{\ph\rho\sigma}\nn\\
&-\frac1{2}\, g_{\mu\lambda}\left[\beta_1\mathbb{1}
+\beta_2e_1\mathbb{1}+\beta_3(e_2\mathbb{1}+S^2)\right]^\lambda_{\ph\lambda\delta}
\delta S^\delta_{\ph\delta\nu}+(\mu\leftrightarrow\nu)\,,
\end{align}
where it is understood that $S$ is given in terms of the background solution $\gmn$ and the chosen $\fmn$ (as opposed to $\delta S^\rho_{\ph\rho\sigma}$ and $\delta [S^2]^\rho_{\ph\rho\sigma}$ which also depends on $h_{\mu \nu}$). In order to arrive at the form above we have manipulated the sums in order to isolate all the terms which involve the explicit variation of the square root matrix $S$ on the last row. All other terms either contain $\hmn$ directly or through $\delta[S^2]^\rho_{\ph\rho\sigma}=-g^{\rho\lambda}h_{\lambda\kappa}[S^2]^{\kappa}_{\ph\kappa\sigma}$.

Note that the fact that $f_{\mu \nu}$ is fixed suggests another way to go for the special case of the $\beta_1$ model. Indeed, in this case, the field equations can be rewritten as 
in Eq.~\eqref{b1_Ssol}. Squaring this expression, we can obtain the following expression for $f_{\mu \nu}$
\ba \label{fmnRicci}
f_{\mu\nu}=\dfrac{\beta_0^2}{9\beta_1^2}g_{\mu\nu}-\dfrac{2\beta_0}{3m^2\beta_1^2}\left(R_{\mu\nu}-\frac{1}{6}R g_{\mu\nu}\right)+\dfrac{1}{m^4\beta_1^2}\left(R_{\mu\rho}R^{\rho}_{\ph\rho\nu}+\frac{1}{36}R^2g_{\mu\nu}-\frac{1}{3}RR_{\mu\nu}\right) \,,
\ea
and then use the fact that $ \delta f_{\mu\nu}=0 $ to obtain linearized field equations by writing the variation of the right hand side above, which is not very difficult to get and in particular much easier than linearizing $S$.  We give in appendix \ref{otherway2} the corresponding expressions and check also there that the two ways to go indeed match, as they should. This requires the use of some non trivial syzygies which are given in appendix \ref{otherway1}.

\subsection{Variation of the square root matrix \& the Cayley-Hamilton theorem}
In order to compute the variation of a square root matrix, $\delta\sqrt{S^2}$, once the variation of the matrix itself, $\delta{S^2}$, is known it is very useful to use the Cayley-Hamilton theorem.\footnote{We thank S.F.~Hassan for initial discussions about this point.} The Cayley-Hamilton theorem is the statement that any square matrix (say $d\times d$) itself obeys its characteristic (or secular) equation, 
 i.e. the polynomial equation obeyed by its eigenvalues $\lambda$ 
\be\label{CH}
p(\lambda)\equiv\det(S-\lambda\mathbb{1})=0\,.
\ee
The theorem therefore states that $p(S)=0$ and can be conveniently stated using the scalar functions $e_k(S)$ as 
\be\label{CH1}
\sum_{n=0}^d(-1)^nS^{d-n}e_n(S)=0\,. 
\ee 
For a $4\times4$ matrix $S$ the Cayley-Hamilton theorem guarantees that,
\be\label{CH_4}
S^4-e_1S^3+e_2S^2-e_3S+e_4\mathbb{1}=0\,.
\ee
An obvious virtue of this theorem is that it can be used to reduce any polynomial in $S$  so that it contains no powers of $S$ higher than three, a fact we make frequent use of in our analysis. 

Computing the variation of (\ref{CH1}), using \eqref{d_e_n}, we get,
\be\label{dSdS2}
\sum_{n=0}^{d-1}(-1)^ne_n(S)\delta[S^{d-n}]
=\frac1{2}\sum_{n=1}^d\sum_{k=1}^n(-1)^{n+k}e_{n-k}(S)S^{d-n}\Tr[S^{k-2}\delta S^2]\,,
\ee
This relation is very useful in computing $\delta S$ since the right hand side is given entirely in terms of $\delta S^2$. We may proceed somewhat further in the general case by splitting the sum on the left hand side into even and odd terms, but since we are mostly interested in dealing with $d=4$ for now we will simply restrict to this case immediately. 
In $d=4$, as can be seen starting from (\ref{CH_4}),  only two terms contain odd powers of the matrix $S$, one of which is linear and the other one cubic. Writing $S^3=S^2\cdot S$ and isolating the terms with $\delta S$ we then obtain the following matrix equation for the variation,\footnote{We note that if we had chosen to write $S^3=S\cdot S^2$ we would have instead obtained an equivalent equation.}
\begin{align}\label{d_S_eq}
[e_3\mathbb{1}+e_1S^2]\delta S=&\,\delta S^2[S^2-e_1S]+[S^2+e_2\mathbb{1}]\delta S^2-\frac1{2}\sum_{n=1}^4\sum_{k=1}^n(-1)^{n+k}e_{n-k}S^{4-n}\Tr[S^{k-2}\delta S^2]\,.
\end{align}
In order to completely isolate $\delta S$ we must assume that the matrix,
\be\label{Xdef}
\mathbb{X}\equiv e_3\mathbb{1}+e_1S^2\,,
\ee
appearing on the left hand side of equation (\ref{d_S_eq}) is invertible, i.e.~that $\det(\mathbb{X})\neq0$. This invertibility is equivalent to the statement that the matrices $S$ and its negative $-S$ have 
no common eigenvalues, as it is proven in appendix  \ref{app_sylv}. This matches the necessary and sufficient condition 
 for the existence of a unique solution of the Sylvester equation (\ref{SylvdS}) obeyed by $\delta S$ (see e.g. \cite{Sylvesterpoly}) and we will assume here that this condition holds. Note that backgrounds for which this is not the case are such that the theory cannot be unambiguously linearized. When $\mathbb{X}$ is invertible, using again the Cayley-Hamilton theorem one can compute the inverse as,
\be\label{Xinv}
\mathbb{X}^{-1}=-\frac{1}{e_4(\mathbb{X})}\sum_{n=0}^3(-1)^n\mathbb{X}^{3-n}e_n(\mathbb{X})\,,
\ee
where we recall that $e_4(\mathbb{X})=\det(\mathbb{X})$. With the assumed invertibility of $\mathbb{X}$ and using \eqref{d_S_eq} we then arrive at the matrix relation,
\be\label{d_S}
\delta S=\,\mathbb{X}^{-1}\left[\delta S^2[S^2-e_1S]+[S^2+e_2\mathbb{1}]\delta S^2
-\frac1{2}\sum_{m=1}^4\sum_{k=1}^m(-1)^{m+k}e_{m-k}S^{4-m}\Tr[S^{k-2}\delta S^2]\right]\,.
\ee
Note that all terms on the right hand side are now given entirely in terms of $\delta S^2$ and can therefore easily be computed in terms of metric perturbations using $S^2=g^{-1}f$. Eq.~\eqref{d_S} thus provides the general expression for the variation of the square root of a matrix in terms of the variation of the matrix itself. We note that this expression might be useful not only for computing perturbations in dRGT theories, but also in their bimetric extension or other generalizations. In the appendix 
\ref{app_sylv} we provide a proof that the above expression (which, as far as we know, is new to this work) matches the one known in the mathematical literature and obtained from the Sylvester equation \cite{Sylvesterpoly}.

\subsection{The general perturbation equations in dRGT theories.}
After having obtained the structure of the generic variation of $\delta S$ we return to the dRGT theories, for which $\delta[S^2]^\rho_{\ph\rho\sigma}=-g^{\rho\lambda}h_{\lambda\kappa}[S^2]^{\kappa}_{\ph\kappa\sigma}$. For our purposes it is convenient to first manipulate the matrix prefactor $\mathbb{X}^{-1}$ (c.f.~Eq.~\eqref{Xinv}) even further. For any $d\times d$ matrix $\mathbb{M}$ we note the identity,
\be
e_n(\mathbb{1}+\lambda\mathbb{M})=\sum_{k=0}^{n}{d-k \choose n-k}\lambda^ke_k(\mathbb{M})\,.
\ee
This can be used to obtain $e_n(\mathbb{X})$ in terms of $e_n(S^2)$. From the definitions \eqref{edef} of the $e_n$ and the Cayley-Hamilton theorem \eqref{CH1} it is then possible to prove that,\footnote{Similar relations were noted in \cite{Guarato:2013gba} in terms of the eigenvalues of triangular matrices but here we provide the completely covariant expressions in terms of the matrices themselves.}
\be
e_1(S^2)=e_1^2-2e_2\,,\quad e_2(S^2)=e_2^2-2e_1e_3+2e_4\,,\quad
e_3(S^2)=e_3^2-2e_2e_4\,,\quad  e_4(S^2)=e_4^2\,.
\ee
These relations can be used to show that, e.g.
\be\label{detX}
e_4(\mathbb{X})=\left[e_1^2e_4+e_3^2-e_1e_2e_3\right]^2\,,
\ee
along with similar relations for the remaining $e_n(\mathbb{X})$. Using these relations together with a binomial expansion of $\mathbb{X}^{3-n}$ and a judicious use of the Cayley-Hamilton relation \eqref{CH_4} to reduce all powers of $S$ higher than three, we find that \eqref{Xinv} can be written,
\be
\mathbb{X}^{-1}=\frac{\left(e_3-e_1e_2\right)\mathbb{1}+e_1^2S-e_1S^2}{e_1^2e_4+e_3^2-e_1e_2e_3}\,.
\ee
Inserting this in \eqref{d_S}, again making use of the Cayley-Hamilton theorem to reduce all powers of $S$ higher than three, we arrive at
\begin{align}\label{d_Sm}
\dfrac{{\delta S}^{\lambda}_{\hphantom{\lambda}\mu}}{\delta g_{\rho\sigma}}=\,\, &
\dfrac{1}{2}\,g^{\nu \lambda} \Bigl[\, e_{4}\,c_{1}\,\Bigl(
\delta_{\nu}^{\rho}\delta^{\sigma}_{\mu}+\delta_{\nu}^{\sigma}\delta_{\mu}^{\rho}-g_{\mu\nu}g^{\rho\sigma}
\Bigr) + e_{4}\,c_{2}\,\Bigl(
S_{\nu}^{\rho}\delta^{\sigma}_{\mu}+S_{\nu}^{\sigma}\delta_{\mu}^{\rho}-S_{\mu\nu}g^{\rho\sigma}
-\gmn S^{\rho\sigma}
\Bigr) \nn\\
&- e_{3}\,c_{1}\,\Bigl(
\delta_{\nu}^{\rho}S^{\sigma}_{\mu}+\delta_{\nu}^{\sigma}S_{\mu}^{\rho} \Bigr) 
+\left(e_{2}\,c_{1}-e_{4}\,c_{3}+e_{3}\,c_{2}\right)\,S_{\mu\nu}S^{\rho\sigma}
\nn\\[1pt]
& +e_{4}\,c_{3}\,\Bigl[\delta^{\sigma}_{\mu}[S^{2}]_{\nu}^{\rho} 
+\delta_{\mu}^{\rho}[S^{2}]_{\nu}^{\sigma}-g^{\rho\sigma}[S^{2}]_{\mu\nu}+\delta_{\nu}^{\rho}[S^{2}]^{\sigma}_{\mu}+\delta_{\nu}^{\sigma}[S^{2}]_{\mu}^{\rho}-g_{\mu\nu}[S^{2}]^{\rho\sigma}
\Bigr] \nn\\[1pt]
& - e_{3}\,c_{2}\,\Bigl(
S_{\nu}^{\rho}S^{\sigma}_{\mu}+S_{\nu}^{\sigma}S_{\mu}^{\rho} \Bigr)
-e_{3}\,c_{3}\,\Bigl(
S^{\sigma}_{\mu}[S^{2}]_{\nu}^{\rho}+S_{\mu}^{\rho}[S^{2}]_{\nu}^{\sigma}+S_{\nu}^{\rho}[S^{2}]^{\sigma}_{\mu}+S_{\nu}^{\sigma}[S^{2}]_{\mu}^{\rho}
\Bigr) \nn\\[1pt]
& +\left(e_{3}\,c_{3}-e_{1}\,c_{1}\right)\,\left(
S^{\rho\sigma}[S^{2}]_{\mu\nu}+S_{\mu\nu}[S^{2}]^{\rho\sigma} \right)
 - \left(c_{1}-e_{2}\,c_{3}\right)\,\Bigl(
[S^{2}]_{\nu}^{\rho}[S^{2}]^{\sigma}_{\mu}+[S^{2}]_{\nu}^{\sigma}[S^{2}]_{\mu}^{\rho}
\Bigr) \nn\\[1pt]
&
+c_{4}\,[S^{2}]_{\mu\nu}[S^{2}]^{\rho\sigma}
+c_{1}\,\Bigl([S^{3}]_{\mu\nu}S^{\rho\sigma}+S_{\mu\nu}[S^{3}]^{\rho\sigma}\Bigr)
+c_{2}\Bigl([S^{3}]_{\mu\nu}[S^{2}]^{\rho\sigma}+[S^{2}]_{\mu\nu}[S^{3}]^{\rho\sigma}\Bigr)\nn\\
&+ c_{3}\,[S^{3}]_{\mu\nu}[S^{3}]^{\rho\sigma} \,\Bigr] \,,
\end{align}
where the coefficients $c_i$ are given by,
\begin{align}
c_{1} &= \dfrac{e_{3}-e_{1}e_{2}}{-e_{1}e_{2}e_{3}+e_{3}^{2}+e_{1}^{2}e_{4}} \,,\;
 c_{2} = \dfrac{e_{1}^{2}}{-e_{1}e_{2}e_{3}+e_{3}^{2}+e_{1}^{2}e_{4}} \,, \;\nn\\
c_{3}& = \dfrac{-e_{1}}{-e_{1}e_{2}e_{3}+e_{3}^{2}+e_{1}^{2}e_{4}} \,,\;
c_{4} = \dfrac{e_{3}-e^3_{1}}{-e_{1}e_{2}e_{3}+e_{3}^{2}+e_{1}^{2}e_{4}} \,,\;
\end{align}
and we recall that all the $e_n$ are functions of $S$ and indices are raised and lowered using the background solution $\gmn$.

Let us summarize our results so far. Inserting \eqref{d_Sm} into \eqref{dVmn} allows us to write the equations of motion \eqref{d_eom1} as,
\be \label{fieldsummary}
\delta E_{\mu\nu} = \delta\mathcal{G}_{\mu\nu}
+m^2\,\mathcal{M}_{\mu\nu}^{\ph\mu\ph\nu\rho\sigma}h_{\rho\sigma}\,.
\ee
where we recall that,
\be
\delta\mathcal{G}_{\mu\nu}=\mathcal{E}_{\mu\nu}^{\ph\mu\ph\nu\rho\sigma}h_{\rho\sigma}
+\tfrac1{2}\left[g_{\mu\nu} R^{\rho\sigma}-\delta^\rho_\mu \delta^\sigma_\nu R\right]h_{\rho\sigma}\,,
\ee
and the mass matrix $\mathcal{M}_{\mu\nu}^{\ph\mu\ph\nu\rho\sigma}$ is defined through,
\be\label{Mdef}
\mathcal{M}_{\mu\nu}^{\ph\mu\ph\nu\rho\sigma}\equiv\frac{\p\Vmn}{\p g_{\rho\sigma}}\,,
\ee
and given explicitly by,
\be\label{Mgen}
\begin{aligned}
\mathcal{M}_{\mu\nu}^{\ph\mu\ph\nu\rho\sigma} = \,&\frac1{2}V_{\mu}^{\ \sigma}\delta_{\nu}^{\rho} - \frac1{2}\left(\beta_2\delta_{\mu}^{\lambda}+\beta_3\left(e_1\delta_{\mu}^{\lambda}-S_{\mu}^{\lambda}\right)\right)[S^2]_{\lambda}^{\sigma}\,\delta_{\nu}^{\rho} \\
& + \frac{1}{4}\,\sum_{n=1}^3\sum_{k=1}^n\sum_{m=1}^k(-1)^{n+k+m}\beta_n\,e_{k-m}
[S^{n-k}]^\lambda_{\ph\lambda\mu}\,g_{\nu\lambda}\,g^{\tau(\rho}\,[S^m]^{\sigma)}_{\ph\sigma\tau} \\
&  - \frac1{2}\left(\beta_1\,\delta^{\tau}_{\ph\tau\lambda}+\beta_2\,e_1\,\delta^{\tau}_{\ph\tau\lambda}+\beta_3\left(e_2\,\delta^{\tau}_{\ph\tau\lambda}+ [S^2]^{\tau}_{\ph\tau\lambda}\right)\right)\dfrac{\delta S^{\lambda}_{\ph\lambda\mu}}{\delta g_{\rho\sigma}}\,g_{\nu\tau} +(\mu\leftrightarrow\nu) \,.
\end{aligned}
\ee
Together with the variation of the square root in \eqref{d_Sm} this can now be used to get the expression of $\delta E_{\mu\nu}$ as an operator acting on the perturbation $\hmn$. This completes the general calculation of the equations of motion for perturbations in massive gravity with an arbitrary reference metric $\fmn$ and an arbitrary background solution $\gmn$. Note that in these equations, $\fmn$ only arises through the presence of the tensor $S_{\mu \nu}$, and one can in fact trade one for the other. Moreover, we recall that in the special case of the $\beta_1$ model introduced above, one can explicitly express $S_{\mu \nu}$ as a function of $g_{\mu \nu}$ and its curvature using equation (\ref{b1_Ssol}), henceby obtaining in this case field equations for the metric perturbation $h_{\mu \nu}$ which only depend on $g_{\mu \nu}$ and its curvature. One can also check that to the lowest non trivial order (i.e. at linear order) in curvature, the field equations found here agree with the ones of \cite{Buchbinder:1999ar} which investigates at this order the consistent coupling of a massive graviton to a curved background.

\label{sec4C}
\subsection{Action for a massive graviton} 
\label{sec4D}
The above obtained equations of motion derive of course from an action obtained from the dRGT action expanded at quadratic order into the metric perturbation. This action, to be given below for completeness, can hence also be used as a starting point for a theory of a massive graviton, with the correct number of propagating polarizations on an arbitrary background (which we prove here in the case of the $\beta_1$ model). 

In order to get the action for perturbations, the interaction term 
\be
S_m = -2M_g^2 m^2\int\td^4 x\sqrt{|g|}V \,.
\ee
can be expanded at quadratic order into the metric perturbations as 
\be \label{dS1}
S_m^{\mathrm{(2,I)}} = -m^2\int\td^4x\delta g^{\mu\nu}\frac{\p^2(\sqrt{|g|}V)}{\p g^{\mu\nu}\p g^{\rho\sigma}}\delta g^{\rho\sigma}\,,
\ee
where,
\be
\frac{\p^2(\sqrt{|g|}V)}{\p g^{\mu\nu}\p g^{\rho\sigma}}=\frac1{2}\sqrt{|g|}\mathcal{M}_{\rho\sigma\mu\nu}
+\frac1{4}\sqrt{|g|}\gmn V_{\rho\sigma}\,,
\ee\newline
or 
\be
S_m^{\mathrm{(2,II)}} = -m^2\int\td^4x\delta g_{\mu\nu}\frac{\p^2(\sqrt{|g|}V)}{\p g_{\mu\nu}\p g_{\rho\sigma}}\delta g_{\rho\sigma}\,,
\ee
where,
\be  \label{dS2}
\frac{\p^2(\sqrt{|g|}V)}{\p g_{\mu\nu}\p g_{\rho\sigma}}=\frac1{2}\sqrt{|g|}\mathcal{M}^{\rho\sigma\mu\nu}
+\frac1{4}\sqrt{|g|}g^{\mu\nu} V^{\rho\sigma}-\frac1{2}\sqrt{|g|}g^{\mu\rho}V^{\nu\sigma}
-\frac1{2}\sqrt{|g|}g^{\mu\sigma}V^{\nu\rho}\,.
\ee\newline
depending on whether one uses $\delta g^{\mu\nu}$ or $\delta g_{\mu\nu}$ as independent variables. 
Note that the left hand side of the above equation (\ref{dS2}) has been computed by Guarato \& Durrer \cite{Guarato:2013gba} who worked it out however in terms of the eigenvalues of an $S$ that was taken to be in upper triangular form.
We have verified that our results coincide with theirs for diagonal $\gmn$ and $\fmn$. We note however that our expressions are explicitly covariant while this is not the case for those of Ref.~\cite{Guarato:2013gba} which uses a coordinate system where $S$ is in triangular form.\footnote{The analysis done in the last part of \cite{Guarato:2013gba} about FLRW space-time is however not correct, as we will discuss in more detail in section \ref{FLRWsection}, since Ref.~\cite{Guarato:2013gba} does not work with valid background solutions.} 

The relations \eqref{dS1} or \eqref{dS2} provide an explicit map between the second variation of the interaction term of the dRGT action and $\mathcal{M}_{\mu\nu}^{\ph\mu\ph\nu\rho\sigma}$ which can be considered as a mass matrix. Note that this matrix depends in general (either implicitly, or explictly in the $\beta_1$ model where $S_{\mu \nu}$ can be traded for $R_{\mu\nu}$) on the curvature of the background metric $g_{\mu \nu}$.
Computing now the full action \eqref{SdRGT} at quadratic order we get easily (using the background field equations as well as the above expressions) the following quadratic action 
\begin{equation}
S^{(2)} = - \frac{1}{2} M_g^2 \int\td^4x\sqrt{|g|} h_{\mu \nu} \left(\tilde{\mathcal{E}}^{\mu\nu\rho\sigma}+ m^2 {\mathcal{M}}^{\mu\nu\rho\sigma}\right) h_{\rho \sigma} \,,
\end{equation}
where the expression for the kinetic tensor $\tilde{\mathcal{E}}^{\mu\nu\rho\sigma}$ has been provided before. This action, when varied, yield indeed the field equations (\ref{fieldsummary}). Note however that neither ${\mathcal{E}}^{\mu\nu\rho\sigma}$ nor ${\mathcal{M}}^{\mu\nu\rho\sigma}$ as given in \eqref{dGmn}, \eqref{EinstOp} and \eqref{Mgen} are manifestly symmetric in the pairwise index interchange $(\mu\nu)\leftrightarrow(\rho\sigma)$. The combination above however is indeed symmetric as can be seen after using the background equations to express everything either in terms of $S$ or in terms of curvature.

\section{The $\beta_1$ model \& the fifth constraint}


The aim of this section is to show, in a covariant and Lagrangian way, that the above derived field equations indeed lead to a scalar constraint according to the general guidelines outlined in Sec.~\ref{sec:outline}. We will do so for the $\beta_1$ model introduced in Sec.~\ref{sec:b1}, where any reference to the non dynamical metric $\fmn$ can be fully eliminated, and provide the details of the calculation for this case.

We recall that the $\beta_1$ model is defined by $\beta_2=\beta_3=0$. In this case the mass matrix $\mathcal{M}_{\mu\nu}^{\ph\mu\ph\nu\rho\sigma}$ is easily obtained from \eqref{Mgen} and the equations of motion are given by,
\be
\delta E_{\mu\nu}=\delta\mathcal{G}_{\mu\nu}+m^2\beta_0\,h_{\mu\nu}+m^2\beta_1\,e_{1}\,h_{\mu\nu} -
m^2\beta_1\,h^{\lambda}_{\ (\nu}\,S_{\mu)\lambda} -
\frac{m^2\beta_1}{2}\,g_{\mu\nu}\,S^{\sigma}_{\ \lambda}\,h^{\lambda}_{\
\sigma} - m^2\beta_1\,\dfrac{g_{\lambda (\nu}\delta S^{\lambda}_{\ph\lambda\mu)}}{\delta
g_{\rho\sigma}}\,h_{\rho\sigma} \,,
\ee
where the last term is obtained from \eqref{d_Sm}. 
As discussed in Sec.~\ref{sec:outline}, we search for a linear combination of the scalars $\Phi_i$ and $\Psi_i$
 (with $i\leq 3$), i.e. for a set of coefficients $\{\ccc_i,\ddd_i\}$, such that  
(\ref{lincombwesearch}) holds.
An explicit calculation of the relevant scalars $\Phi_i$ reveals that
\begin{subequations}
\ba
\Phi_{0} &\sim& -\,\Bigl(\nabla_{\rho}\nabla_{\sigma}\,h^{\rho\sigma}-\square\,h\Bigr) \,, \\
\Phi_{1} &\sim&
\dfrac{1}{2}\,\Bigl[2\,S^{\mu\nu}\,\nabla_{\alpha}\nabla_{\mu}\,h^{\alpha}_{\nu} -
S^{\mu\nu}\,\square\,h_{\mu\nu} - S^{\mu\nu}\,\nabla_{\nu}\nabla_{\mu}\,h +
e_1\Phi_{0}\Bigr] \,, \\
\Phi_{2} &\sim&
\dfrac{1}{2}\,\Bigl[2\,[S^2]^{\mu\nu}\,\nabla_{\alpha}\nabla_{\mu}\,h^{\alpha}_{\nu}
- [S^2]^{\mu\nu}\,\square\,h_{\mu\nu} - [S^2]^{\mu\nu}\,\nabla_{\nu}\nabla_{\mu}\,h
+ \left(e_1^2-2e_2\right)\Phi_{0}\Bigr] \,,
\\
\Phi_{3} &\sim&
\dfrac{1}{2}\,\Bigl[2\,[S^3]^{\mu\nu}\,\nabla_{\alpha}\nabla_{\mu}\,h^{\alpha}_{\nu}
- [S^3]^{\mu\nu}\,\square\,h_{\mu\nu} - [S^3]^{\mu\nu}\,\nabla_{\nu}\nabla_{\mu}\,h
+ \left(e_1^3-3e_1e_2+3e_3\right)\Phi_{0}\Bigr] \,,
\ea
\end{subequations}
where $h=h^\rho_{\ph\rho\rho}$. In order to compute the $\Psi_i$ efficiently we first note that the variation of the Bianchi identity, i.e.~$\delta\left(\nabla^{\mu}\,\mathcal{G}_{\mu\nu}\right)=0$, implies,
\be \label{Bianpert}
\nabla^{\mu}\,\delta \mathcal{G}_{\mu\nu}-h^{\mu\rho}\,\nabla_{\rho}\,\mathcal{G}_{\mu\nu}-\mathcal{G}_{\sigma\nu}\,\nabla_{\rho}\,h^{\sigma\rho}+\frac{1}{2}\,\mathcal{G}_{\sigma\nu}\,\nabla^{\sigma}\,h
-\frac{1}{2}\,\mathcal{G}_{\mu\sigma}\,\nabla_{\nu}\,h^{\mu\sigma} = 0 \,.
\ee
Using Eq.~(\ref{beta1field}) (or equivalently (\ref{b1_Ssol})) to express above every occurence of the Einstein tensor $\mathcal{G}_{\mu\nu}$ in terms if $S_{\mu \nu}$, we then get the following expression for the $\Psi_i$'s, where again we only retain terms with two derivatives acting on $h_{\mu \nu}$,
\ba
\begin{aligned}
\Psi_{i}\sim\,\frac{m^2}{2}\Bigl[& \mathcal{M}_{\mu\nu}^{\ \ \rho\sigma}\,[S^i]^{\nu}_{\
\lambda}\,\nabla^{\lambda}\nabla^{\mu}\,h_{\rho\sigma} +
\beta_{1}\,[S^{i+1}]_{\sigma\lambda}\,\nabla^{\lambda}\nabla_{\rho}\,h^{\rho\sigma} -
\dfrac{\beta_{1}}{2}\,[S^{i+1}]_{\sigma\lambda}\,\nabla^{\lambda}\nabla^{\sigma}\,h \\
& + \dfrac{\beta_{1}}{2}\,S_{\sigma\mu}\,[S^i]^{\nu}_{\
\lambda}\,\nabla^{\lambda}\nabla_{\nu}\,h^{\mu\sigma} -
\left(\beta_{0}+\beta_{1}\,e_1\right)\,[S^i]^{\nu}_{\
\lambda}\,\nabla^{\lambda}\nabla_{\rho}\,h^{\rho}_{\nu} \Bigr] \,.
\end{aligned}
\ea
We stress here that we can view Eq.~(\ref{beta1field}) (or equivalently (\ref{b1_Ssol})) just as a definition of $S$ in terms of the curvatures of the metric $\gmn$ and that using this equation does not result in any restriction on the metric $\gmn$.
A close examination of the expressions above reveals that the $\Phi_i$'s and the $\Psi_i$'s are linear combinations of 26 different 
irreducible scalar terms, $\aleph_i$, which we denote in the following way (see footnote \ref{footnotelist})
\begin{align}\label{scalars1}
& J_1=\Jaa \,, &&J_2=\Jbb \,,&& \\
& I_1=\Iaa \,,\, &&I_2=\Ibb \,, &&I_3=\Icc \,, \\
& H_1=\Haa \,, &&H_2=\Hbb \,, &&H_3=\Hcc \,, \\
& G_1=\Gaa \,, &&G_2=\Gbb \,,&& \\
& F_1=\Faa \,, &&F_2=\Fbb \,, &&F_3=\Fcc \,, \\
& E_1=\Eaa \,, &&E_2=\Ebb \,, &&E_3=\Ecc \,, \\
& D_1=\Daa \,, &&D_2=\Dbb \,,&& \\
& C_1=\Caa \,, &&C_2=\Cbb \,, &&C_3=\Ccc \,, \\
& B_1=\Baa \,, &&B_2=\Bbbb \,, &&B_3=\Bcc \,, \\
& A_1=\Aaa \,, &&A_2=\Abb \, .&&
\label{scalars2}
\end{align}
For future reference, we order the $\aleph_i$ from left to right and top to bottom, i.e. we define $\aleph_1 = J_1$, $\aleph_2=J_2$, $\aleph_3= I_1$, $\aleph_4=I_2,$ {\it etc}. The expression of the $\Phi_i$'s in terms of these are given by
\ba \label{defphi0}
\Phi_{0} &\sim& -\,\Bigl(J_1-J_2\Bigr) \,, \\
\Phi_{1} &\sim&
\dfrac{1}{2}\,\Bigl[2\,I_1 -
I_3- I_2 -
e_1\left(J_1-J_2\right)\Bigr] \,, \\
\Phi_{2} &\sim&
\dfrac{1}{2}\,\Bigl[2\,H_1
- H_3-H_2
-\left(e_1^2-2e_2\right)\left(J_1-J_2\right)\Bigr] \,,
\\
\Phi_{3} &\sim& \label{defphi3}
\dfrac{1}{2}\,\Bigl[2\,F_1
-F_3-F_2
-\left(e_1^3-3e_1e_2+3e_3\right)\left(J_1-J_2\right)\Bigr] \,,
\ea
where we have used that $I_1\sim S^{\mu\nu}\,\nabla_{\alpha}\nabla_{\mu}\,h^{\alpha}_{\nu}$ after commuting derivatives together with similar relations for $H_1$ and $F_1$.
The expressions for the $\Psi_i$ in terms of the scalars $\aleph_i$ are given by
\begin{align}
\Psi_0 =&\, \tfrac{m^2\beta_1}{2} \left[ -c_1e_4\left(\Ja-\tfrac{1}{2}\Jb\right) +\left(c_1e_3-c_2e_4\right)\Ia -\tfrac{1}{2}c_1e_3\,\Ib +\tfrac{1}{2}c_2e_4\,\Ic \right.\nn\\
&\left.-c_3e_4\left(2\,\Ha-\tfrac{1}{2}\Hb-\tfrac{1}{2}\Hc\right)
+c_2e_3\,\Ga-\tfrac{1}{2}\left(c_1e_2+c_2e_3-c_3e_4\right)\,\Gb+2c_3e_3\,\Ea \right.\nn\\
&\left.+\tfrac{1}{2}\left(c_1e_1-c_3e_3\right)\left(\Eb+\Ec\right)+\left(c_1-c_3e_2\right)\Da\right. \nn\\
& \left. +\tfrac{1}{2}\left(c_2e_1-c_1+c_3e_2\right)\Db-\tfrac{1}{2}c_1\left(\Cb+\Cc\right) -\tfrac{1}{2}c_2\left(\Bb+\Bc\right) -\tfrac{1}{2}c_3\,\Ab \right] \,, 
\label{defpsi0}
\end{align}
\begin{align}
 \Psi_1 =&\, \tfrac{m^2\beta_1}{2} \left[ -c_1e_4\left(\Ia-\tfrac{1}{2}\Ib-\tfrac{1}{2}\Ic\right) +c_1e_3\left(\Ha-\tfrac{1}{2}\Hb\right)+\tfrac{1}{2}c_2e_4\,\Hc-c_2e_4\,\Ga \right.\nn\\
&\left.-\tfrac{1}{2}\left(c_1e_3-c_2e_4\right)\Gb 
-c_3e_4\left(\Fa-\tfrac{1}{2}\Fb-\tfrac{1}{2}\Fc\right) \right.\nn\\
&\left.+\left(c_2e_3-c_3e_4\right)\left(\Ea-\tfrac{1}{2}\Eb-\tfrac{1}{2}\Ec\right) +c_3e_3\left(\Da-\Db\right) \right. \nonumber \\
& \left. +c_3e_3\left(\Ca-\tfrac{1}{2}\Cb-\tfrac{1}{2}\Cc\right) +\left(c_1-c_3e_2\right)\left(\Ba-\tfrac{1}{2}\Bb-\tfrac{1}{2}\Bc\right) \right] \,, 
\end{align}
\ba
\Psi_2 &=& \tfrac{m^2\beta_1}{2} \left[ c_3e_4^2\left(\Ja-\tfrac{1}{2}\Jb\right) -c_3e_3e_4\left(2\Ia-\tfrac{1}{2}\Ib-\tfrac{1}{2}\Ic\right) -\left(c_1-c_3e_2\right)e_4\left(2\Ha-\tfrac{1}{2}\Hb-\tfrac{1}{2}\Hc\right) \right. \nonumber \\
&& \left. +c_3e_3^2\,\Ga+\tfrac{1}{2}\left(c_1e_4-c_3e_3^2\right)\,\Gb +\left(c_1e_3+c_2e_4\right)\left(\Fa-\tfrac{1}{2}\Fb\right)+\left(c_1e_3-2c_3e_2e_3-c_2e_4\right)\Ea \right. \nonumber  \\
&& -\tfrac{1}{2}\left(c_1e_3-c_3e_2e_3-c_2e_4\right)\left(\Eb+\Ec\right)-\left(c_1e_2-c_3e_2^2+c_3e_4\right)\Da
\nn \\
&& -\tfrac{1}{2}\left(c_2e_3-c_1e_2+c_3e_2^2-c_3e_4\right)\Db \nonumber \\
&&  \left. +\tfrac{1}{2}c_3e_4\left(\Cb+\Cc\right) -\tfrac{1}{2}c_3e_3\left(\Bb+\Bc\right) -\tfrac{1}{2}\left(c_1-c_3e_2\right)\Ab \right]  \,, 
\ea
\ba
\Psi_3 &=& \tfrac{m^2\beta_1}{2} \left[ -\left(c_1e_3e_4+c_2e_4^2\right)\left(\Ja-\tfrac{1}{2}\Jb\right) +\left(c_1e_3^2+c_2e_3e_4+c_3e_4^2\right)\left(\Ia-\tfrac{1}{2}\Ib\right) -\tfrac{1}{2}c_3e_4^2\,\Ic \right. \nonumber \\
&& \left. -\left(c_1e_2e_3+c_2e_2e_4+c_3e_3e_4\right)\left(\Ha-\tfrac{1}{2}\Hb\right)+\tfrac{1}{2}c_3e_3e_4\,\Hc -c_3e_3e_4\left(\Ga-\Gb\right) \right. \nonumber \\
&& \left. +\left(c_1e_1e_3-c_1e_4+c_2e_1e_4+c_3e_2e_4\right)\left(\Fa-\tfrac{1}{2}\Fb\right)+\tfrac{1}{2}\left(c_1-c_3e_2\right)e_4\,\Fc \right. \nonumber \\
&&   \left.-\left(c_1e_4-c_3e_3^2-c_3e_2e_4\right)\left(\Ea-\tfrac{1}{2}\Eb-\tfrac{1}{2}\Ec\right)+\left(c_1-c_3e_2\right)e_3\,\Da \right. \nn \\
&& -\tfrac{1}{2}\left(c_1e_3-c_2e_4-2c_3e_2e_3\right)\Db  \nonumber\\
&&   \left.-\left(c_3e_2e_3+c_2e_4\right)\Ca
 -\tfrac{1}{2}\left(c_1-c_3e_2\right)e_3\left(\Cb+\Cc\right)
\right.\nn\\
&&\left.
 -\left(c_1e_2-c_3e_2^2+c_3e_4\right)\left(\Ba-\tfrac{1}{2}\Bb-\tfrac{1}{2}\Bc\right) \right] \,.
\label{defpsi3}
\ea
Using the above expressions for the $\Phi_i$ and the $\Psi_i$ one can obtain the explicit form of the coefficients $\alpha_i$ as defined in Eq.~(\ref{linalphacoef}). The corresponding expressions of the $\alphacoef_i$ are given in appendix \ref{listalpha}. Having obtained these expressions, it may seem now that solving equation (\ref{lincombwesearch}) requires to solve 26 equations for 7 unknown ratios of the $\{\ccc_i,\ddd_i\}$, these 26 equations being obtained by setting to zero independently each scalar coefficient of each $\aleph_i$ appearing in the linear combination on the left hand side of (\ref{lincombwesearch}). A closer examination of these equations shows that they reduce to a homogeneous system of 8 independent equations whose only solution is the trivial one (with vanishing $u_i$ and $v_i$). However, one notices that the $\aleph_i$ of equations~\eqref{scalars1} to \eqref{scalars2} are not all independent from each other but are linked by non trivial identities - syzygies - as we now show. These identities, as a consequence of the "second fundamental theorem" of invariant theory \cite{Procesi,Sneddon}, can be derived using the Cayley-Hamilton theorem which we already stated before. For clarity, let us rewrite here this theorem for an arbitrary $4\times 4$ matrix $M$, as
\be \label{CHM}
M^{4} = e_{1}(M)  M^{3}-  e_{2}(M)M^{2} +  e_{3}(M) M - e_{4}(M)\mathbb{1}\,.
\ee
One can then apply this to a matrix $M$ built out of 4 arbitrary matrices $A,B,C,D$ and four arbitrary real numbers $\{x_i\}$ in the form, 
\be \label{defMatM}
M=x_0 A+x_1 B+x_2C+x_3D\,.
\ee
Now, because the $\{x_i\}$ as well as the matrices $A,B,C,D$ are arbitrary, it means that in equation (\ref{CHM}) the terms which have the same degree of homogeneity in the $\{x_i\}$ must each yield separate identities between the matrices $A,B,C,D$. Once these identities are obtained, one can replace in them $A$ by $h$, $B$ by $S$, $C$ by $S^2$ and $D$ by $S^3$ to get non trivial matrix syzygies, denoted here as  $[{\cal I}_{k}]^\mu_{\hphantom{\mu}\nu}=0$, between the tensors of interest. 
For our purpose here, since we are interested in relations where $\hmn$ appears linearly, it is enough to consider syzygies derived by looking at terms linear in $x_0$ (but this is the only restriction imposed on the $x_i$). The corresponding syzygies $[{\cal I}_{k}]^\mu_{\hphantom{\mu}\nu}=0$ are gathered in appendix \ref{Syzygies}.
The next step is to contract each of these syzygies with $\nabla_\mu \nabla^\nu$ and retain there only the terms with two derivatives acting directly on the perturbation $\hmn$.
In the whole process, we use again the Cayley-Hamilton theorem to eliminate any powers of $S$ greater than three. 
At the end of the process we obtain the following four independent identities from the corresponding equations (\ref{syz1})-(\ref{syz4}),
\ba \label{SY1}
&& \begin{aligned} & \Aa-\Ab-e_2\left(\Da-\Db\right) -e_3\left(2\Fa-\Fb-\Fc\right)-\left(e_4-e_1e_3\right)\left(\Ga-\Gb\right) \\
& -\left(e_4-e_1e_3\right)\left(2\Ha-\Hb-\Hc\right)-e_2e_3\left(2\Ia-\Ib-\Ic\right) +e_3^2\left(\Ja-\Jb\right) \sim 0 \,, \end{aligned} \\[10pt]
&& 2\Ba-\Bb-\Bc -e_1\left(\Da-\Db\right)+e_3\left(\Ga-\Gb\right) -e_4\left(2\Ia-\Ib-\Ic\right) \sim 0 \,, \\[10pt]
&& \begin{aligned} & 2\Ca-\Cb-\Cc +\left(\Da-\Db\right)+e_1\left(2\Fa-\Fb-\Fc\right)-\left(e_1^2-e_2\right)\left(\Ga-\Gb\right) \\
& -e_1^2\left(2\Ha-\Hb-\Hc\right) +e_1e_2\left(2\Ia-\Ib-\Ic\right)-\left(e_1e_3+e_4\right)\left(\Ja-\Jb\right) \sim 0 \,, \end{aligned} \\[10pt]
&& \begin{aligned} & 2\Ea-\Eb-\Ec +\left(2\Fa-\Fb-\Fc\right) -e_1\left(\Ga-\Gb\right)-e_1\left(2\Ha-\Hb-\Hc\right) \\
& +e_2\left(2\Ia-\Ib-\Ic\right)-e_3\left(\Ja-\Jb\right) \sim 0 \,. \end{aligned} \label{SY2}
\ea
Fortunately, these equations are precisely enough to change our original system of 8 (which only have the trivial solution $\ccc_i=\ddd_i=0$) equations into the following system of 7 equations,
\ba
&& e_1\,\ddd_0-e_3\,\ddd_2 = 0 \,,\\[5pt]
&& \ddd_0+e_3\,\ddd_3 = 0 \,,\\[5pt]
&& \ddd_1-e_2\,\ddd_3 = 0 \,,\\[5pt]
&& \ccc_0+\frac{m^2\,\beta_1}{4}\,e_4\,\ddd_3 = 0 \,,\\[5pt]
&& \ccc_1 = \ccc_2 = \ccc_3 = 0 \,.
\ea
We recall that we are only interested in 7 ratios of the $\{\ccc_i,\ddd_i\}$, so this system has a unique solution. With this solution we find the identity (\ref{finalidentity}) which in turns results on shell 
into the announced constraint 
\begin{equation} \label{finalconscons}
-\dfrac{m^2\,\beta_1\,e_4}{4}\,\Phi_0 - e_3\,\Psi_0 + e_2\,\Psi_1-e_1\,\Psi_2 + \Psi_3 = 0 \,.
\end{equation}
We will below detail some applications of our results. In particular, looking at the flat space-time case will allow us to prove that the above constraint (\ref{finalconscons}) is independent of the vector constraints (\ref{vectorconsgen}) as it should in order to remove an extra degree of freedom. 

\section{Some applications}

\label{Applic}
\subsection{Flat space-time limit}
We first look at the flat space-time limit of the above constraint (\ref{finalconscons}). Choosing $\gmn$ to be the canonical Minkowski metric $\eta_{\mu \nu}$ we get from  
(\ref{b1_Ssol}) that in this case, one has 
\ba \label{flatS}
S^\rho_{~\nu}= - \frac{\beta_0}{3 \beta_1}\delta^\rho_{\nu}\;.
\ea
And one obtains the field equations operator
\ba 
\delta \bar{E}_{\mu \nu} & \equiv &  \delta \bar{\cal G}_{\mu \nu}  -\frac{m^2 \beta_0}{6} \left(h_{\mu \nu} - h \; \eta _{\mu \nu}\right) \,,
\ea
which corresponds to (\ref{FlatFE1}) with the identification 
$\barm^2 = - m^2 \beta_0/3$. This of course implies that one should impose $\beta_0<0$ in order to have a healthy massive mode. Furthermore, because $e_k(c \cdot \mathbb{1}) =  c^k\,{4 \choose k}$ one finds from (\ref{flatS}) that 
\ba \label{enflat}
e_1 = -\frac{4 \beta_0}{3 \beta_1}\,, \;\;\, e_2 = \frac{ 2 \beta_0^2}{3 \beta_1^2}\,, \; \;\, e_3 = -\frac{4 \beta_0^3}{27 \beta_1^3}\,, \;\;\, e_4 = \frac{\beta_0^4}{81\beta_1^4}\,,
\ea
and in this case the constraint (\ref{finalconscons}) reads 
\ba
\frac{\beta^3_0}{54 \beta_1^3}\left(\partial^\mu \partial^\nu \delta \bar{E}_{\mu \nu} - m^2 \frac{\beta_0}{6} \eta^{\mu \nu} \delta \bar{E}_{\mu \nu}\right) = -m^4 \frac{\beta_0^5}{648 \beta_1^3}\,h  = 0 \,,
\ea
where the first equality expresses the identity (\ref{finalidentity}) (which holds here as stated above, i.e. without the neglecting of terms $\sim 0$) and the second one results from the field equation. Hence, ones finds back 
(\ref{consMin1}) and (\ref{consMin2}), as expected. Moreover, this shows that the constraint (\ref{finalconscons}) is in the general case (i.e. not assuming anymore that $\gmn$ is flat) independent of the vector constraints (\ref{vectorconsgen}).

\subsection{Einstein space-times}
Let us study now the more general case of Einstein space-time backgrounds. Such a space-time obeys the following equations
\ba
R_{\mu \nu} &=& - {\cal G}_{\mu \nu} = \Lambdabak g_{\mu \nu} \\
R &=& 4 \Lambdabak\,,
\ea
where $\Lambdabak$ stands here for the background cosmological constant.
Inserting these relation into (\ref{b1_Ssol}) we get 
\ba \label{Einsteinspace}
S^\rho_{~\nu}= - \frac{\tilde{\beta}_0}{3 \beta_1}\delta^\rho_{\nu}\,,
\ea
where $\tilde{\beta}_0$ is given by 
\ba
\tilde{\beta}_0 = \beta_0 - \frac{\Lambdabak}{m^2}\,.
\ea
This means in particular (see equation (\ref{flatS})) that the relations (\ref{enflat}) hold provided one replaces there $\beta_0$ by $\tilde{\beta}_0$. 
It is easy to use the above relation (\ref{Einsteinspace}) to deduce from the flat space-time discussion above that the field equations operator is given by 
\ba 
\delta {E}_{\mu \nu} & \equiv &  \delta {\cal G}_{\mu \nu} +\Lambdabak h_{\mu \nu}  -\frac{m^2 \tilde{\beta}_0}{6} \left(h_{\mu \nu} - h \; g_{\mu \nu}\right)\,.
\ea
This can of course also be checked by a direct calculation from Eqs.~(\ref{fieldsummary})-(\ref{Mgen}). 
Notice that the first two terms on the right hand side are just the linear expansion of ${\cal G}_{\mu \nu} + \Lambdabak g_{\mu \nu}$ while the last two terms have the flat space-time Fierz-Pauli form provided one replaces $\beta_0$ by $\tilde{\beta_0}$. This structure of the massive spin-2 equations is well known and similarly to the flat space case the mass of the spin-2 field is identified by $m_{\mathrm{FP}}^2=-m^2\tilde\beta_0/3$. A direct calculation, or the use again of the previous results for flat space-time (together with the identity (\ref{Bianpert}) which now contains a non trivial piece with undifferentiated $h_{\mu  \nu}$), yields the following form for the constraint (\ref{finalconscons})
\ba
\frac{\tilde{\beta}^3_0}{54 \beta_1^3}\left(\nabla^\mu \nabla^\nu \delta E_{\mu \nu} - m^2 \frac{\tilde{\beta}_0}{6} g^{\mu \nu} \delta E_{\mu \nu}\right) = -m^4 \frac{\tilde{\beta}_0^5}{648 \beta_1^3}\, h \left(1+\frac{2 \Lambdabak}{\tilde{\beta}_0 m^2}\right)  = 0\,.
\ea
Apart from the obvious general solution $h=0$ there is now also a special case when the parameters satisfy $2\Lambda=-\tilde\beta_0m^2$ and the constraint equation is automatically satisfied and becomes an identity. This corresponds to the well known Higuchi bound $2\Lambda=3m_{\mathrm{FP}}^2$ \cite{Higuchi:1986py} and for this special case it is known that the theory has an enhanced linear gauge symmetry which renders the helicity-0 mode of the massive spin-2 field non-propagating \cite{Deser:2001us}. This is however an artifact of linearization and it has been shown that the symmetry only exists for Einstein backgrounds and can not be extended to general backgrounds within the framework of dRGT theories \cite{deRham:2013wv, Deser:2013uy, Fasiello:2013woa, Joung:2014aba}.

\subsection{FLRW backgrounds} \label{FLRWsection}
We now assume that $g_{\mu\nu}$ parametrizes a Friedmann-Lema\^{\i}tre-Robertson-Walker space-time in conformal time, with the line element given by 
\be \label{gFLRW}
\md s_{g}^2=g_{\mu\nu}\md x^{\mu}\md x^{\nu} = a^2(\eta)\left[-\md\eta^2+\gamma_{ij}\md x^{i}\md x^{j}\right] \,.
\ee
The corresponding non-vanishing components of $S$ are obtained using \eqref{b1_Ssol} as 
\ba
&& S_{0}^{0} = -\dfrac{\beta_0}{3\beta_1}+\dfrac{1}{m^2\beta_1\,a^2}\left(-\mathcal{H}^2+2\mathcal{H}'-K\right)\,, \\
&& S^{i}_{j} = \left[-\dfrac{\beta_0}{3\beta_1}+\dfrac{1}{m^2\beta_1\,a^2}\left(\mathcal{H}^2+K\right)\right]\delta^{i}_{j} \,,
\ea
all other components being zero, where $K=0,\pm 1$ parametrizes the spatial curvature as usual, and $\mathcal{H} = a'/a$ is the Hubble factor of the conformal metric (\ref{gFLRW}) (a prime denoting here a derivative with respect to the conformal time $\eta$). For convenience we further define
\ba
&& s_0 = -\dfrac{\beta_0}{3\beta_1}+\dfrac{1}{m^2\beta_1\,a^2}\left(-\mathcal{H}^2+2\mathcal{H}'-K\right) \,, \\
&& s_1 = -\dfrac{\beta_0}{3\beta_1}+\dfrac{1}{m^2\beta_1\,a^2}\left(\mathcal{H}^2+K\right) \,,
\ea
such that $S_{0}^{0} = s_0$ and $S^{i}_{j} = s_1\delta^{i}_{j}$. One also has the following useful relations 
\ba \label{relationseeee}
e_1=s_0+3s_1 \,, \qquad e_2=3(s_0+s_1)s_1 \,, \qquad e_3=(3s_0+s_1)s_1^2 \,, \qquad e_4=s_0 s_1^3 \,,
\ea
Squaring the tensor $S$ we obtain $f_{\mu\nu}$ as 
\begin{align} \label{f1FLRW}
f_{00} &= -a^2\,s_0^2 = -a^2\left[\dfrac{\beta_0^2}{9\beta_1^2}+\dfrac{2\beta_0}{3m^2\beta_1^2\,a^2}\left(\mathcal{H}^2-2\mathcal{H}'+K\right)+\dfrac{1}{m^4\beta_1^2\,a^4}\left(\mathcal{H}^2-2\mathcal{H}'+K\right)^2\right]\,, \\
f_{ij} &= a^2\,s_1^2\,\gamma_{ij} = a^2\left[ \dfrac{\beta_0^2}{9\beta_1^2}-\dfrac{2\beta_0}{3m^2\beta_1^2\,a^2}\left(\mathcal{H}^2+K\right)+\dfrac{1}{m^4\beta_1^2\,a^4}\left(\mathcal{H}^2+K\right)^2 \right]\gamma_{ij} \,.
\label{f2FLRW}
\end{align}
We can then compute the mass matrix,
\begin{align}
\mathcal{M}_{00}^{\ph 0\ph 000} &= \beta_0 + 3s_1\beta_1 
= \frac{3}{m^2\,a^2}\left(\mathcal{H}^2+K\right) \,, \\
\mathcal{M}_{00}^{\ph 0\ph 0ij} &= \frac{1}{2}\beta_1\,s_1\,\gamma^{ij} 
= \left[-\frac{1}{6}\beta_0 +\frac{1}{2m^2\,a^2}\left(\mathcal{H}^2+K\right)\right]\gamma^{ij} \,, \\
\mathcal{M}_{ij}^{\ph i\ph j00} &= \frac{1}{2}\beta_1\,s_0\,\gamma_{ij} 
= \left[-\frac{1}{6}\beta_0 -\frac{1}{2m^2\,a^2}\left(\mathcal{H}^2+K-2\mathcal{H}'\right)\right]\gamma_{ij} \,,\\
\mathcal{M}_{0i}^{\ph 0\ph i0j} &= \frac{1}{2}\left[\beta_0+\beta_1\left(\dfrac{s_0^2+3s_0s_1+3s_1^2}{s_0+s_1}\right)\right]\delta^{i}_{j} \nn\\
&= \left[\dfrac{1}{2m^2\,a^2}\left(\mathcal{H}^2+2\mathcal{H}'+K\right) +\dfrac{\dfrac{1}{9}m^2a^2\beta_0^2-\dfrac{2}{3}\beta_0\left(\mathcal{H}^2+K\right)+\dfrac{1}{m^2a^2}\left(\mathcal{H}^2+K\right)^2}{4\mathcal{H}'-\dfrac{4}{3}\beta_0 m^2a^2} \right]\delta^{i}_{j} \,, \\
\mathcal{M}_{ij}^{\ph i\ph jkl} &= \left[\beta_0+\frac{1}{2}\beta_1\left(2s_0+5s_1\right)\right]\delta_{(i}^{k}\delta_{j)}^{l} - \frac{1}{2}\beta_1\,s_1\,\gamma_{ij}\gamma^{kl} \nn\\
& = \left[-\frac{1}{6}\beta_0 +\frac{1}{2m^2\,a^2}\left(3\left(\mathcal{H}^2+K\right) +4\mathcal{H}'\right)\right]\delta^{k}_{(i}\delta^{l}_{j)} +\left[\frac{1}{6}\beta_0-\frac{1}{2m^2\,a^2}\left(\mathcal{H}^2+K\right)\right]\gamma_{ij}\gamma^{kl} \,.
\end{align}
Using the above expressions in those of sections \ref{sec4C} and \ref{sec4D} we obtain the equations of motion and action for a sound (as far as the counting of d.o.f. is concerned) massive graviton propagating in an arbitrary FLRW space-time. 
Notice that a similar problem was also studied in Ref. \cite{Guarato:2013gba} starting from a similar point as us. However, it was assumed there that both the dynamical metric $g_{\mu \nu}$ and the non dynamical metric $f_{\mu \nu}$ can be taken to be conformally flat. Here we have shown that this assumption is not correct. Indeed, assuming (\ref{gFLRW}) leads to the form (\ref{f1FLRW})-(\ref{f2FLRW}) for  $f_{\mu \nu}$. Said in another way, this means that the metrics considered in \cite{Guarato:2013gba} are not solutions of the background field equations of the dRGT model, and it can in fact be checked that any such solutions where  $g_{\mu \nu}$ and $f_{\mu \nu}$ are proportional to each other via a coefficient $c$ must be such that $c$ is a constant. As a result, the theory found in \cite{Guarato:2013gba} for a massive graviton on FLRW space-time is not the correct one obtained from expanding the dRGT model and hence very likely does not yield the correct number of propagating degrees of freedom. In constrast, in our case, one can check using in particular the relations (\ref{relationseeee}), that the equation \eqref{finalconscons} is indeed a constraint.

\section{Conclusion}
In this paper we have  given the detailed derivation of the field equations for a massive graviton propagating on an arbitrary background metric as obtained from the dRGT model. For the simplest of these models, we have shown how the non dynamical metric $f_{\mu \nu}$ which appears in the formulation of the dRGT theories can be explicitly eliminated out of these equations leaving just one metric $g_{\mu \nu}$ which serves as a background. Starting from these equations, we have shown how to obtain a scalar constraint which plays there the role of the tracelessness of the massive Fierz-Pauli graviton obtained on flat backgrounds. We stress that, once one considers the linearized equations as expressed uniquely in terms of  $g_{\mu \nu}$ and its perturbation $h_{\mu \nu}$, the constraint is identically satisfied for an arbitrary $g_{\mu \nu}$, i.e. irrespectively of the fact it is a solution of the background equations of motion. Hence, the obtained linear equations (respectively the obtained action) for $h_{\mu \nu}$ can be considered as giving general equations of motion (respectively a general action) for a massive graviton with (at most) 5 polarizations on an arbitrary background, irrespectively from its origin in the dRGT model. It involves in particular a mass matrix which depends in an explicit and very specific way on the background curvature. This work can be extended in various directions. E.g. an obvious question is whether a similar constraint can be obtained when one does not assume that $\beta_2$ and $\beta_3$ vanish, i.e. in the general dRGT model \cite{USPREP}.
Note further that massive gravity, including its dRGT version, suffers from various possible pathologies introduced in particular in the reviews \cite{deRham:2014zqa,Hinterbichler:2011tt}. Some of these pathologies, which exact meaning and consequences are subject to a debate, would also arise in the linearized version of the theory we presented here. This concerns in particular the issue of superluminal propagation and related causality issues discussed e.g. in \cite{Burrage:2011cr,Burrage:2012ja,Gruzinov:2011sq,Deser:2013qza,Deser:2013eua,Deser:2012qx,Yu:2013owa,Deser:2014hga,Deser:2015wta}, and which can be seen at the level of the linearized theory and might invalidate some specific background. We refer the reader to these latter references as well as to the reviews \cite{deRham:2014zqa,Hinterbichler:2011tt} for further discussion of these and other potential problems of massive gravity.

{\bf Acknowledgments:} The researches of CD and MvS leading to these results have
received funding from the European Research Council under the European
Community’s Seventh Framework Programme (FP7/2007-2013 Grant Agreement
no. 307934).
In the process of checking our calculations, we have heavily used the \textit{xTensor} package~\cite{xTensor}
developed by J.-M.~Mart\'{\i}n-Garc\'{\i}a for
\textit{Mathematica}. We thank Angnis Schmidt-May, Fawad Hassan and Kurt Hinterbichler for discussions.

\appendix

\section{An inductive proof of the variation $\delta e_n$}\label{deltaeninduc}
Here, for completeness, we provide an inductive proof for \eqref{d_e_n}. Starting from $e_0=1$ and,
\be\label{app_edef}
e_n(S)=-\frac{1}{n}\sum_{k=1}^{n}(-1)^k\Tr[S^k]\,e_{n-k}(S)\,,\qquad n\geq1\,,
\ee
the claim is that
\be\label{app_d_e_n}
\delta e_n(S)=-\sum_{k=1}^n(-1)^k\Tr[S^{k-1}\delta S]\,e_{n-k}(S)\,,\qquad n\geq1\,.
\ee
For $n=1$ we find that this formula gives $\delta e_1(S)=\Tr[\delta S]$, which is the correct result since $e_1(S)=\Tr[S]$. Let us now assume that \eqref{app_d_e_n} is true for all $k\leq n$ and compute the variation of $\delta e_{n+1}$ directly from the defninition \eqref{app_edef},
\begin{align}
\delta e_{n+1}(S)&=-\frac1{n+1}\sum_{k=1}^{n+1}(-1)^k\left[\Tr[\delta S^k]e_{n+1-k}+\Tr[S^k]\delta e_{n+1-k}\right]\nn\\
&=\frac{1}{n+1}\sum_{k=1}^{n+1}\sum_{m=1}^{n+1-k}(-1)^{k+m}e_{n+1-k-m}\left[\frac1{n+1-k}\Tr[\delta S^k]\Tr[S^m]
+\frac1{m}\Tr[\delta S^m]\Tr[S^k]\right]\nn\\
&=\frac{1}{n+1}\sum_{k=1}^{n+1}\sum_{m=1}^{n+1-k}(-1)^{k+m}e_{n+1-k-m}\Tr[\delta S^k]\Tr[S^m]\left[\frac1{n+1-k}
+\frac1{k}\right]\nn\\
&=\sum_{k=1}^{n+1}(-1)^k\Tr[S^{k-1}\delta S]\frac1{n+1-k}
\sum_{m=1}^{n+1-k}(-1)^{m}\Tr[S^m]e_{n+1-k-m}\nn\\
&=-\sum_{k=1}^{n+1}(-1)^k\Tr[S^{k-1}\delta S]e_{n+1-k}\,.
\end{align}
To get to the second line we used \eqref{app_edef} and \eqref{app_d_e_n} together with the identity $m\Tr[S^{m-1}\delta S]=\Tr[\delta S^m]$. To get to the third line we
used that for an arbitrary function $f(k,m)$ one has 
$\sum_{k=1}^{n+1}\sum_{m=1}^{n+1-k}f(k,m)=\sum_{k=1}^{n+1}\sum_{m=1}^{n+1-k}f(m,k)$. We then simply combined everything and again used \eqref{app_edef}.

\section{The Sylvester equation and another form for the square root variation}\label{app_sylv}
A much studied matrix equation in mathematics is the Sylvester equation,
\be
AX-XB=C\,,
\ee
where $A$, $B$ and $C$ are known square matrices and $X$ is to be determined. The matrices $A$ and $B$ are matrices not necessarily of the same rank. The Sylvester theorem \cite{sylvesterThm} states that for every matrix $C$, there is a unique solution for $X$ if and only if $\sigma(A)\cap\sigma(B)=\emptyset$ where $\sigma(M)$ denotes the spectrum of $M$. In this case the solution for $X$ is also a known polynomial in the matrices $A$, $B$ and $C$ given by \cite{Sylvesterpoly}
\be
X=q_B^{-1}(A)\sum_{k=1}^{r_B}\sum_{m=0}^{k-1}(-1)^ke_{4-k}(B)A^{k-m-1}CB^m\,,
\ee
where $r_B$ is the rank of the matrix $B$ and $q_B(M)$ is the characteristic polynomial for $B$, i.e.
\be
q_B(M)=\sum_{n=0}^d(-1)^ne_n(B)M^{d-n}\,.
\ee
The connection to our work can be seen when considering equation (\ref{SylvdS}) which shows that $\delta S$ obeys a Sylvester equation 
with the identifications $A=S$, $B=-S$ and $C=\delta S^2=\delta \gf$. The Sylvester theorem then guarantees that there is a unique solution for $\delta  S$ if and only if $\sigma(S)\cap\sigma(-S)=\emptyset$, i.e.~when $S$ and $-S$ share no common eigenvalues. The solution for $\delta S$ can then be written,
\be \label{B4}
\delta S=\frac1{2}\X^{-1}\sum_{k=1}^4\sum_{m=0}^{k-1}(-1)^me_{4-k}S^{k-m-2}\delta S^2S^m\,,
\ee
where $\X=e_3\mathbb{1}+e_1S^2$ as defined in \eqref{Xdef}. The converse statement that $S$ and $-S$ have at least one common eigenvalue, i.e.~$\sigma(S)\cap\sigma(-S)\neq\emptyset$ is in fact equivalent to the statement that $\X$ is not invertible as we will now demonstrate explicitly. For this, consider $\lambda\in\sigma(S)$, which means that it solves the characteristic equation of $S$,
\be\label{lambda_eq1}
\lambda^4-e_1\lambda^3+e_2\lambda^2-e_3\lambda+e_4=0\,.
\ee
If in addition $\lambda\in\sigma(-S)$ we must also have that it solves the characteristic equation of $-S$,
\be
\lambda^4+e_1\lambda^3+e_2\lambda^2+e_3\lambda+e_4=0\,,
\ee
or equivalently, using \eqref{lambda_eq1},
\be
e_1\lambda^3+e_3\lambda=0\,.
\ee
The solution $\lambda=0$ implies that $e_4=\det S=0$, i.e.~that $S$ is not invertible which is a basic assumption in our setup so the second equation reduce to,
\be
e_3+e_1\lambda^2=0\,.
\ee
The set of equations for $\lambda$ can now be written,
\be
\lambda^4+e_2\lambda^2+e_4=0\,,\qquad e_3+e_1\lambda^2=0\,.
\ee
One solution of the second equation is given by $e_1=e_3=0$. In this case it is obvious that $\X=e_1+e_3S^2$ is not invertible since it is the zero matrix. The other solution is $\lambda^2=-e_3/e_1$ which result in the condition,
\be
e_3^2-e_1e_2e_3+e_1^2e_4=0\,.
\ee
From Eq.~\eqref{detX} this condition can be seen to imply $e_4(\X)=\det\X=0$, again implying that $\X$ is not invertible so we have shown that $\sigma(S)\cap\sigma(-S)\neq\emptyset$ implies that $e_4(\X)=0$. Conversely, from Eq.~\eqref{detX} it is clear that $e_4(\X)=0$ implies,
\be
e_3^2-e_1e_2e_3+e_1^2e_4=0\,.
\ee
Obviously, by the assumption that $S$ is invertible we have that $e_4\neq0$. Hence in this equation $e_1=0\Leftrightarrow e_3=0$ from which it follows that any $\lambda\in\sigma(S)$ is also in $\sigma(-S)$. For $e_1\neq0$ we define $\lambda^2=-e_3/e_1$ and it is then again clear that this $\lambda$ will be in both $\sigma(S)$ and $\sigma(-S)$. We have thus established the equivalence between the statements $e_4(\X)=0$ and $\sigma(S)\cap\sigma(-S)\neq\emptyset$.


We also explicitly verify below that the solution of the Sylvester equation for $\delta S$ given by (\ref{B4}) 
coincides with the expression for $\delta S$ given in \eqref{d_S} and that we derived earlier. We denote $\delta S_{\vert S}$ the expression  (\ref{B4}) and still use $\delta S$ for the expression \eqref{d_S}. In order to simplify the expressions, we look at the difference $2S\X\left(\delta S_{\vert S}-\delta S\right)S$ and obtain that (where we use notations introduced in appendix \ref{Syzygies})
\ba\label{diffS-M}
\left[2S\X\left(\delta S_{\vert S}-\delta S\right)S\right]^{\mu}_{\nu}&=&-e_4\left([G'_1]^{\mu}_{\ph\mu\nu}-[G'_2]^{\mu}_{\ph\mu\nu}+2[H'_1]^{\mu}_{\ph\mu\nu}
-[H'_2]^{\mu}_{\ph\mu\nu}-[H'_3]^{\mu}_{\ph\mu\nu}\right) \nn\\
&&+e_3\left(2[E'_1]^{\mu}_{\ph\mu\nu}
-[E'_2]^{\mu}_{\ph\mu\nu}-[E'_3]^{\mu}_{\ph\mu\nu}\right) \nonumber \\
&&-e_2\left([D'_1]^{\mu}_{\ph\mu\nu}-[D'_2]^{\mu}_{\ph\mu\nu}\right)
+\left([A'_1]^{\mu}_{\ph\mu\nu}-[A'_2]^{\mu}_{\ph\mu\nu}\right) \,,
\ea
The difference~\eqref{diffS-M} can now be rewritten as a function of equations given in appendix \ref{Syzygies},
\be
\left[2S\X\left(\delta S_{\vert S}-\delta S\right)S\right]^{\mu}_{\nu} = \eqref{syz1}+e_3\ \eqref{syz4} \,  = 0\,.
\ee
As we assume $\X$ and $S$ to be invertible this implies,
\be\label{otherway1} \delta S_{\vert S} = \delta S \,. \ee

\section{Expression of the coefficients $\alphacoef_i$}\label{App2}\label{listalpha}
In this appendix, we give the list of all the coefficients $\alpha_i$ as defined from Eq.~(\ref{linalphacoef}) and computed using the expressions (\ref{defphi0})-(\ref{defphi3}) and (\ref{defpsi0})-(\ref{defpsi3}) given in the main text.
\ba
&& \begin{aligned} \alphacoef_{1} = &\quad -\ccc_0-\tfrac{1}{2}e_1\,\ccc_1-\tfrac{1}{2}\left(e_1^2-2\,e_2\right)\ccc_2-\tfrac{1}{2}\left(e_1^3-3e_1e_2+3e_3\right)\ccc_3\ \nonumber \\
&+\frac{m^2\beta_1}{2} \left[-c_1e_4\,\ddd_0 +c_3e_4^2\,\ddd_2 -\left(c_1e_3e_4+c_2e_4^2\right)\ddd_3\right]  \end{aligned}\nonumber\\
&& \begin{aligned} \alphacoef_{2} = &\quad \ccc_0+\tfrac{1}{2}e_1\,\ccc_1+\tfrac{1}{2}\left(e_1^2-2\,e_2\right)\ccc_2+\tfrac{1}{2}\left(e_1^3-3e_1e_2+3e_3\right)\ccc_3\ \nonumber\\
&+\frac{m^2\beta_1}{4} \left[c_1e_4\,\ddd_0 - c_3e_4^2\,\ddd_2 +\left(c_1e_3e_4+c_2e_4^2\right)\ddd_3\right]  \end{aligned}\nonumber\\
&&\begin{aligned} \alphacoef_{3} = &\quad \ccc_1+\frac{m^2\beta_1}{2} \left[\left(c_1e_3-c_2e_4\right)\ddd_0 -c_1e_4\,\ddd_1 -2c_3e_3e_4\,\ddd_2 +\left(c_1e_3^2+c_2e_3e_4+c_3e_4^2\right)\ddd_3\right]  \end{aligned}\nonumber\\
&&\begin{aligned} \alphacoef_{4} = &\quad -\frac{1}{2}\ccc_1+\frac{m^2\beta_1}{4} \left[-c_1e_3\,\ddd_0 +c_1e_4\,\ddd_1 +c_3e_3e_4\,\ddd_2 -\left(c_1e_3^2+c_2e_3e_4+c_3e_4^2\right)\ddd_3\right]  \end{aligned}\nonumber\\
&&\begin{aligned} \alphacoef_{5} = &\quad -\frac{1}{2}\ccc_1+\frac{m^2\beta_1}{4} \left[c_2e_4\,\ddd_0 +c_1e_4\,\ddd_1 +c_3e_3e_4\,\ddd_2 -c_3e_4^2\,\ddd_3\right]  \end{aligned}\nonumber\\
&&\begin{aligned} \alphacoef_{6} = &\quad \ccc_2+\frac{m^2\beta_1}{2} \left[-2c_3e_4\,\ddd_0 +c_1e_3\,\ddd_1 -2\left(c_1-c_3e_2\right)e_4\,\ddd_2 -\left(c_1e_2e_3+c_2e_2e_4+c_3e_3e_4\right)\ddd_3\right]  \end{aligned}\nonumber\\
&&\begin{aligned} \alphacoef_{7} = &\quad -\frac{1}{2}\ccc_2+\frac{m^2\beta_1}{4} \left[ c_3e_4\,\ddd_0 -c_1e_3\,\ddd_1 +\left(c_1-c_3e_2\right)e_4\,\ddd_2 +\left(c_1e_2e_3+c_2e_2e_4+c_3e_3e_4\right)\ddd_3\right]  \end{aligned}\nonumber\\
&&\begin{aligned} \alphacoef_{8} = &\quad -\frac{1}{2}\ccc_2+\frac{m^2\beta_1}{4} \left[ c_3e_4\,\ddd_0 +c_2e_4\,\ddd_1 +\left(c_1-c_3e_2\right)e_4\,\ddd_2 +c_3e_3e_4\,\ddd_3\right]  \end{aligned}\nonumber\\
&&\begin{aligned} \alphacoef_{9} = &\quad \frac{m^2\beta_1}{2} \left[ c_2e_3\,\ddd_0 -c_2e_4\,\ddd_1 +c_3e_3^2\,\ddd_2 -c_3e_3e_4\,\ddd_3\right]  \end{aligned}\nonumber\\
&&\begin{aligned} \alphacoef_{10} = &\quad \frac{m^2\beta_1}{4} \left[ -\left(c_1e_2+c_2e_3-c_3e_4\right)\ddd_0 -\left(c_1e_3-c_2e_4\right)\ddd_1 +\left(c_1e_4-c_3e_3^2\right)\ddd_2 +2c_3e_3e_4\,\ddd_3\right]  \end{aligned}\nonumber\\
&&\begin{aligned} \alphacoef_{11} = &\quad \ccc_3+\frac{m^2\beta_1}{2} \left[-c_3e_4\,\ddd_1 +\left(c_1e_3+c_2e_4\right)\ddd_2 +\left(c_1e_1e_3-c_1e_4+c_2e_1e_4+c_3e_2e_4\right)\ddd_3\right]  \end{aligned}\nonumber\\
&&\begin{aligned} \alphacoef_{12} = &\quad -\frac{1}{2}\ccc_3+\frac{m^2\beta_1}{4} \left[c_3e_4\,\ddd_1 -\left(c_1e_3+c_2e_4\right)\ddd_2 -\left(c_1e_1e_3-c_1e_4+c_2e_1e_4+c_3e_2e_4\right)\ddd_3\right]  \end{aligned}\nonumber\\
&&\begin{aligned} \alphacoef_{13} = &\quad -\frac{1}{2}\ccc_3+\frac{m^2\beta_1}{4} \left[c_3e_4\,\ddd_1 +\left(c_1e_4-c_3e_2e_4\right)\ddd_3\right]  \end{aligned}\nonumber\\
&&\begin{aligned} \alphacoef_{14} = &\quad \frac{m^2\beta_1}{2} \left[2c_3e_3\,\ddd_0 +\left(c_2e_3-c_3e_4\right)\ddd_1 +\left(c_1e_3-2c_3e_2e_3-c_2e_4\right)\ddd_2 \right.\nn\\
&\qquad\qquad\left.-\left(c_1e_4-c_3e_3^2-c_3e_2e_4\right)\ddd_3\right]  \end{aligned}\nonumber\\
&&\begin{aligned} \alphacoef_{15} = &\quad \frac{m^2\beta_1}{4} \left[\left(c_1e_1-c_3e_3\right)\ddd_0 -\left(c_2e_3-c_3e_4\right)\ddd_1 -\left(c_1e_3-c_3e_2e_3-c_2e_4\right)\ddd_2 \right.\nn\\
&\qquad\qquad\left.+\left(c_1e_4-c_3e_3^2-c_3e_2e_4\right)\ddd_3\right]  \end{aligned}\nonumber\\
&&\begin{aligned} \alphacoef_{16} = &\quad \frac{m^2\beta_1}{4} \left[\left(c_1e_1-c_3e_3\right)\ddd_0 -\left(c_2e_3-c_3e_4\right)\ddd_1 -\left(c_1e_3-c_3e_2e_3-c_2e_4\right)\ddd_2 \right.\nn\\
&\qquad\qquad\left.+\left(c_1e_4-c_3e_3^2-c_3e_2e_4\right)\ddd_3\right]  \end{aligned}\nonumber
\ea
\ba
&&\begin{aligned} \alphacoef_{17} = &\quad \frac{m^2\beta_1}{2} \left[\left(c_1-c_3e_2\right)\ddd_0 +c_3e_3\,\ddd_1 -\left(c_1e_2-c_3e_2^2+c_3e_4\right)\ddd_2 +\left(c_1e_3-c_3e_2e_3\right)\ddd_3\right]  \end{aligned}\nonumber\\
&&\begin{aligned} \alphacoef_{18} = &\quad \frac{m^2\beta_1}{4} \left[\left(c_2e_1-c_1+c_3e_2\right)\ddd_0 -2c_3e_3\,\ddd_1 -\left(c_2e_3-c_1e_2+c_3e_2^2-c_3e_4\right)\ddd_2\right.\nn\\ &\qquad\qquad\left.-\left(c_1e_3-c_2e_4-2c_3e_2e_3\right)\ddd_3\right]  \end{aligned}\nonumber\\
&&\begin{aligned} \alphacoef_{19} = &\quad \frac{m^2\beta_1}{2} \left[c_3e_3\,\ddd_1-\left(c_3e_2e_3+c_2e_4\right)\ddd_3\right]  \end{aligned}\nonumber\\
&&\begin{aligned} \alphacoef_{20} = &\quad \frac{m^2\beta_1}{4} \left[-c_1\,\ddd_0-c_3e_3\,\ddd_1+c_3e_4\,\ddd_2-\left(c_1e_3-c_3e_2e_3\right)\ddd_3\right]  \end{aligned}\nonumber\\
&&\begin{aligned} \alphacoef_{21} = &\quad \frac{m^2\beta_1}{4} \left[-c_1\,\ddd_0-c_3e_3\,\ddd_1+c_3e_4\,\ddd_2-\left(c_1e_3-c_3e_2e_3\right)\ddd_3\right]  \end{aligned}\nonumber \\
&&\begin{aligned} \alphacoef_{22} = &\quad \frac{m^2\beta_1}{2} \left[\left(c_1-c_3e_2\right)\ddd_1-\left(c_1e_2-c_3e_2^2+c_3e_4\right)\ddd_3\right]  \end{aligned}\nonumber\\
&&\begin{aligned} \alphacoef_{23} = &\quad \frac{m^2\beta_1}{4} \left[-c_2\ddd_0-\left(c_1-c_3e_2\right)\ddd_1-c_3e_3\ddd_2+\left(c_1e_2-c_3e_2^2+c_3e_4\right)\ddd_3\right]  \end{aligned}\nonumber \\
&&\begin{aligned} \alphacoef_{24} = &\quad \frac{m^2\beta_1}{4} \left[-c_2\ddd_0-\left(c_1-c_3e_2\right)\ddd_1-c_3e_3\ddd_2+\left(c_1e_2-c_3e_2^2+c_3e_4\right)\ddd_3\right]  \end{aligned}\nonumber\\
&&\begin{aligned} \alphacoef_{25} = &\quad 0  \end{aligned}\nonumber\\
&&\begin{aligned} \alphacoef_{26} = &\quad -\frac{m^2\beta_1}{4} \left[c_3\ddd_0 +\left(c_1-c_3e_2\right)\ddd_2\right]  \end{aligned}\nonumber
\ea

\section{Syzygies linear in $\hmn$}
\label{Syzygies}
Here we gather the syzygies which we obtained as explained below equation (\ref{defMatM}) which are linear in $\hmn$ or its second derivatives. 
First, we define the following 26 tensors
\begin{align}\
& [J'_1]^{\mu\nu}=\Jamn \,, &&[J'_2]^{\mu\nu}=\Jbmn \,,&& \\
& [I'_1]^{\mu\nu}=\Iamn \,, &&[I'_2]^{\mu\nu}=\Ibmn \,, &&[I'_3]^{\mu\nu}=\Icmn \,, \\
& [H'_1]^{\mu\nu}=\Hamn \,, &&[H'_2]^{\mu\nu}=\Hbmn \,, &&[H'_3]^{\mu\nu}=\Hcmn \,, \\
& [G'_1]^{\mu\nu}=\Gamn \,, &&[G'_2]^{\mu\nu}=\Gbmn \,,&& \\
& [F'_1]^{\mu\nu}=\Famn \,, &&[F'_2]^{\mu\nu}=\Fbmn \,, &&[F'_3]^{\mu\nu}=\Fcmn \,, \\
& [E'_1]^{\mu\nu}=\Eamn \,, &&[E'_2]^{\mu\nu}=\Ebmn \,, &&[E'_3]^{\mu\nu}=\Ecmn \,, \\
& [D'_1]^{\mu\nu}=\Damn \,, &&[D'_2]^{\mu\nu}=\Dbmn \,,&& \\
& [C'_1]^{\mu\nu}=\Camn \,, &&[C'_2]^{\mu\nu}=\Cbmn \,, &&[C'_3]^{\mu\nu}=\Ccmn \,, \\
& [B'_1]^{\mu\nu}=\Bamn \,, &&[B'_2]^{\mu\nu}=\Bbmn \,, &&[B'_3]^{\mu\nu}=\Bcmn \,, \\
& [A'_1]^{\mu\nu}=\Aamn \,, &&[A'_2]^{\mu\nu}=\Abmn \,,&&
\end{align}
from which the scalars (\ref{scalars1})-(\ref{scalars2}) are obtained by contraction with $\nabla_\mu \nabla_\nu$ and letting the derivative only hit $h_{\rho \sigma}$.
Using the method explained below Eq.~(\ref{defMatM}), we extract the following 10 syzygies from the Cayley-Hamilton relation, using the matrix (\ref{defMatM}) (with the replacement $[A]^{\mu}_{\ph\mu\nu}=[h]^{\mu}_{\ph\mu\nu}$, $[B]^{\mu}_{\ph\mu\nu}=[S]^{\mu}_{\ph\mu\nu}$, $[C]^{\mu}_{\ph\mu\nu}=[S^2]^{\mu}_{\ph\mu\nu}$ and $[D]^{\mu}_{\ph\mu\nu}=[S^3]^{\mu}_{\ph\mu\nu}$) and keeping only the terms linear in $x_0$.  \\
From $ x_0\,x_1\,x_2\,x_3 $ we get
\begin{align}
& 4\left([A'_1]^{\mu}_{\ph\mu\nu}-[A'_2]^{\mu}_{\ph\mu\nu}\right)+e_1\left(2\,[B'_1]^{\mu}_{\ph\mu\nu}-[B'_2]^{\mu}_{\ph\mu\nu}-[B'_3]^{\mu}_{\ph\mu\nu}\right)+e_1^2\left(2\,[C'_1]^{\mu}_{\ph\mu\nu}-[C'_2]^{\mu}_{\ph\mu\nu}-[C'_3]^{\mu}_{\ph\mu\nu}\right) \nn\\ &-4e_2\left([D'_1]^{\mu}_{\ph\mu\nu}-[D'_2]^{\mu}_{\ph\mu\nu}\right)
 +\left(-e_1^3+e_1e_2+e_3\right)\left(2\,[E'_1]^{\mu}_{\ph\mu\nu}-[E'_2]^{\mu}_{\ph\mu\nu}-[E'_3]^{\mu}_{\ph\mu\nu}\right) \nn\\
&+\left(-3e_3+e_1e_2\right)\left(2\,[F'_1]^{\mu}_{\ph\mu\nu}-[F'_2]^{\mu}_{\ph\mu\nu}-[F'_3]^{\mu}_{\ph\mu\nu}\right)
  +\left(-4e_4+4e_1e_3\right)\left([G'_1]^{\mu}_{\ph\mu\nu}-[G'_2]^{\mu}_{\ph\mu\nu}\right) \nn\\ &+\left(-4e_4+3e_1e_3-e_1^2e_2\right)\left(2\,[H'_1]^{\mu}_{\ph\mu\nu}-[H'_2]^{\mu}_{\ph\mu\nu}-[H'_3]^{\mu}_{\ph\mu\nu}\right) \nn\\ 
& +\left(-e_1e_4-3e_2e_3+e_1e_2^2\right)\left(2\,[I'_1]^{\mu}_{\ph\mu\nu}-[I'_2]^{\mu}_{\ph\mu\nu}-[I'_3]^{\mu}_{\ph\mu\nu}\right) \nn\\
& +\left(-e_1^2e_4+3e_3^2-e_1e_2e_3\right)\left([J'_1]^{\mu}_{\ph\mu\nu}-[J'_2]^{\mu}_{\ph\mu\nu}\right) = 0 \,.\nn
\end{align}
From $ x_0\,x_2^{\ph 2 3} $ we get
\begin{align}
& e_1\left(2\,[B'_1]^{\mu}_{\ph\mu\nu}-[B'_2]^{\mu}_{\ph\mu\nu}-[B'_3]^{\mu}_{\ph\mu\nu}\right)-e_1^2\left([D'_1]^{\mu}_{\ph\mu\nu}-[D'_2]^{\mu}_{\ph\mu\nu}\right)
+e_3\left(2\,[E'_1]^{\mu}_{\ph\mu\nu}-[E'_2]^{\mu}_{\ph\mu\nu}-[E'_3]^{\mu}_{\ph\mu\nu}\right) \nn\\
&+e_3\left(2\,[F'_1]^{\mu}_{\ph\mu\nu}-[F'_2]^{\mu}_{\ph\mu\nu}-[F'_3]^{\mu}_{\ph\mu\nu}\right) 
-e_1e_3\left(2\,[H'_1]^{\mu}_{\ph\mu\nu}-[H'_2]^{\mu}_{\ph\mu\nu}-[H'_3]^{\mu}_{\ph\mu\nu}\right) \nn\\
&+\left(-e_1e_4+e_2e_3\right)\left(2\,[I'_1]^{\mu}_{\ph\mu\nu}-[I'_2]^{\mu}_{\ph\mu\nu}-[I'_3]^{\mu}_{\ph\mu\nu}\right) -e_3^2\left([J'_1]^{\mu}_{\ph\mu\nu}-[J'_2]^{\mu}_{\ph\mu\nu}\right) = 0 \,.\nn
\end{align}
From $ x_0\,x_1^{\ph 1 2}\,x_3 $ we get
\begin{align}
& \left(2\,[B'_1]^{\mu}_{\ph\mu\nu}-[B'_2]^{\mu}_{\ph\mu\nu}-[B'_3]^{\mu}_{\ph\mu\nu}\right)+e_1\left(2\,[C'_1]^{\mu}_{\ph\mu\nu}-[C'_2]^{\mu}_{\ph\mu\nu}-[C'_3]^{\mu}_{\ph\mu\nu}\right) -2e_2\left(2\,[E'_1]^{\mu}_{\ph\mu\nu}-[E'_2]^{\mu}_{\ph\mu\nu}-[E'_3]^{\mu}_{\ph\mu\nu}\right) \nn\\ 
& +\left(-2e_2+e_1^2\right)\left(2\,[F'_1]^{\mu}_{\ph\mu\nu}-[F'_2]^{\mu}_{\ph\mu\nu}-[F'_3]^{\mu}_{\ph\mu\nu}\right) +\left(e_3+3e_1e_2-e_1^3\right)\left([G'_1]^{\mu}_{\ph\mu\nu}-[G'_2]^{\mu}_{\ph\mu\nu}\right) \nn\\ 
& +\left(2e_1e_2-e_1^3\right)\left(2\,[H'_1]^{\mu}_{\ph\mu\nu}-[H'_2]^{\mu}_{\ph\mu\nu}-[H'_3]^{\mu}_{\ph\mu\nu}\right) +\left(e_1^2e_2-2e_2^2-e_4\right)\left(2\,[I'_1]^{\mu}_{\ph\mu\nu}-[I'_2]^{\mu}_{\ph\mu\nu}-[I'_3]^{\mu}_{\ph\mu\nu}\right) \nn\\ 
& +\left(-e_1^2e_3+2e_2e_3-e_1e_4\right)\left([J'_1]^{\mu}_{\ph\mu\nu}-[J'_2]^{\mu}_{\ph\mu\nu}\right) = 0 \,.\nn
\end{align}
From $ x_0\,x_1\,x_2^{\ph 2 2} $ we get
\begin{align}
& 2\left(2\,[B'_1]^{\mu}_{\ph\mu\nu}-[B'_2]^{\mu}_{\ph\mu\nu}-[B'_3]^{\mu}_{\ph\mu\nu}\right)+e_1\left(2\,[C'_1]^{\mu}_{\ph\mu\nu}-[C'_2]^{\mu}_{\ph\mu\nu}-[C'_3]^{\mu}_{\ph\mu\nu}\right) -e_1\left([D'_1]^{\mu}_{\ph\mu\nu}-[D'_2]^{\mu}_{\ph\mu\nu}\right) \nn\\ 
& +\left(-e_1^2+e_2\right)\left(2\,[E'_1]^{\mu}_{\ph\mu\nu}-[E'_2]^{\mu}_{\ph\mu\nu}-[E'_3]^{\mu}_{\ph\mu\nu}\right) +e_2\left(2\,[F'_1]^{\mu}_{\ph\mu\nu}-[F'_2]^{\mu}_{\ph\mu\nu}-[F'_3]^{\mu}_{\ph\mu\nu}\right) \nn\\ &+2e_3\left([G'_1]^{\mu}_{\ph\mu\nu}-[G'_2]^{\mu}_{\ph\mu\nu}\right)
 -e_1e_2\left(2\,[H'_1]^{\mu}_{\ph\mu\nu}-[H'_2]^{\mu}_{\ph\mu\nu}-[H'_3]^{\mu}_{\ph\mu\nu}\right) \nn\\
&+\left(-2e_4+e_2^2\right)\left(2\,[I'_1]^{\mu}_{\ph\mu\nu}-[I'_2]^{\mu}_{\ph\mu\nu}-[I'_3]^{\mu}_{\ph\mu\nu}\right) 
+\left(-e_1e_4-e_2e_3\right)\left([J'_1]^{\mu}_{\ph\mu\nu}-[J'_2]^{\mu}_{\ph\mu\nu}\right) = 0 \,.\nn
\end{align}
From $ x_0\,x_1^{\ph 1 2}\,x_2 $ we get
\begin{align}
& 2\left(2\,[C'_1]^{\mu}_{\ph\mu\nu}-[C'_2]^{\mu}_{\ph\mu\nu}-[C'_3]^{\mu}_{\ph\mu\nu}\right) +2\left([D'_1]^{\mu}_{\ph\mu\nu}-[D'_2]^{\mu}_{\ph\mu\nu}\right) -e_1\left(2\,[E'_1]^{\mu}_{\ph\mu\nu}-[E'_2]^{\mu}_{\ph\mu\nu}-[E'_3]^{\mu}_{\ph\mu\nu}\right) \nn\\
&+e_1\left(2\,[F'_1]^{\mu}_{\ph\mu\nu}-[F'_2]^{\mu}_{\ph\mu\nu}-[F'_3]^{\mu}_{\ph\mu\nu}\right)
+\left(-e_1^2+2e_2\right)\left([G'_1]^{\mu}_{\ph\mu\nu}-[G'_2]^{\mu}_{\ph\mu\nu}\right) \nn\\
&-e_1^2\left(2\,[H'_1]^{\mu}_{\ph\mu\nu}-[H'_2]^{\mu}_{\ph\mu\nu}-[H'_3]^{\mu}_{\ph\mu\nu}\right)
+e_1e_2\left(2\,[I'_1]^{\mu}_{\ph\mu\nu}-[I'_2]^{\mu}_{\ph\mu\nu}-[I'_3]^{\mu}_{\ph\mu\nu}\right)  \nn\\
& +\left(-2e_4-e_1e_3\right)\left([J'_1]^{\mu}_{\ph\mu\nu}-[J'_2]^{\mu}_{\ph\mu\nu}\right) = 0 \,.\nn
\end{align} 
From $ x_0\,x_1^{\ph 1 3} $ we get
\begin{align}
& \left(2\,[E'_1]^{\mu}_{\ph\mu\nu}-[E'_2]^{\mu}_{\ph\mu\nu}-[E'_3]^{\mu}_{\ph\mu\nu}\right)+\left(2\,[F'_1]^{\mu}_{\ph\mu\nu}-[F'_2]^{\mu}_{\ph\mu\nu}-[F'_3]^{\mu}_{\ph\mu\nu}\right)-e_1\left([G'_1]^{\mu}_{\ph\mu\nu}-[G'_2]^{\mu}_{\ph\mu\nu}\right) \nn\\
&-e_1\left(2\,[H'_1]^{\mu}_{\ph\mu\nu}-[H'_2]^{\mu}_{\ph\mu\nu}-[H'_3]^{\mu}_{\ph\mu\nu}\right) 
+e_2\left(2\,[I'_1]^{\mu}_{\ph\mu\nu}-[I'_2]^{\mu}_{\ph\mu\nu}-[I'_3]^{\mu}_{\ph\mu\nu}\right) -e_3\left([J'_1]^{\mu}_{\ph\mu\nu}-[J'_2]^{\mu}_{\ph\mu\nu}\right) = 0 \,.\nn
\end{align}
From $ x_0\,x_1\,x_3^{\ph 3 2} $ we get
\begin{align}
& 3e_1\left([A'_1]^{\mu}_{\ph\mu\nu}-[A'_2]^{\mu}_{\ph\mu\nu}\right)-2e_2\left(2\,[B'_1]^{\mu}_{\ph\mu\nu}-[B'_2]^{\mu}_{\ph\mu\nu}-[B'_3]^{\mu}_{\ph\mu\nu}\right)+e_1e_2\left(2\,[C'_1]^{\mu}_{\ph\mu\nu}-[C'_2]^{\mu}_{\ph\mu\nu}-[C'_3]^{\mu}_{\ph\mu\nu}\right) \nn\\ 
& +\left(-e_4+e_1e_3+e_2^2-e_1^2e_2\right)\left(2\,[E'_1]^{\mu}_{\ph\mu\nu}-[E'_2]^{\mu}_{\ph\mu\nu}-[E'_3]^{\mu}_{\ph\mu\nu}\right) \nn\\
&+\left(-e_4-2e_1e_3+e_2^2\right)\left(2\,[F'_1]^{\mu}_{\ph\mu\nu}-[F'_2]^{\mu}_{\ph\mu\nu}-[F'_3]^{\mu}_{\ph\mu\nu}\right) \nn\\ 
& +\left(-2e_1e_4-2e_2e_3+2e_1^2e_3\right)\left([G'_1]^{\mu}_{\ph\mu\nu}-[G'_2]^{\mu}_{\ph\mu\nu}\right)\nn\\
&+\left(-2e_1e_4+2e_1^2e_3-e_1e_2^2\right)\left(2\,[H'_1]^{\mu}_{\ph\mu\nu}-[H'_2]^{\mu}_{\ph\mu\nu}-[H'_3]^{\mu}_{\ph\mu\nu}\right)  \nn\\ 
& +\left(e_2e_4-2e_1e_2e_3+e_2^3\right)\left(2\,[I'_1]^{\mu}_{\ph\mu\nu}-[I'_2]^{\mu}_{\ph\mu\nu}-[I'_3]^{\mu}_{\ph\mu\nu}\right) \nn\\
&+\left(e_3e_4-e_1e_2e_4
+2e_1e_3^2-e_2^2e_3\right)\left([J'_1]^{\mu}_{\ph\mu\nu}-[J'_2]^{\mu}_{\ph\mu\nu}\right) = 0 \,.\nn
\end{align}
From $ x_0\,x_2^{\ph 2 2}\,x_3 $ we get
\begin{align}
& 2e_1\left([A'_1]^{\mu}_{\ph\mu\nu}-[A'_2]^{\mu}_{\ph\mu\nu}\right)+\left(e_1^2-e_2\right)\left(2\,[B'_1]^{\mu}_{\ph\mu\nu}-[B'_2]^{\mu}_{\ph\mu\nu}-[B'_3]^{\mu}_{\ph\mu\nu}\right)+e_3\left(2\,[C'_1]^{\mu}_{\ph\mu\nu}-[C'_2]^{\mu}_{\ph\mu\nu}-[C'_3]^{\mu}_{\ph\mu\nu}\right)  \nn\\ 
& +\left(e_3-e_1e_2-e_1^3\right)\left([D'_1]^{\mu}_{\ph\mu\nu}-[D'_2]^{\mu}_{\ph\mu\nu}\right) +\left(2e_1e_3-2e_4\right)\left(2\,[E'_1]^{\mu}_{\ph\mu\nu}-[E'_2]^{\mu}_{\ph\mu\nu}-[E'_3]^{\mu}_{\ph\mu\nu}\right) \nn\\ 
& +\left(-2e_4+e_1e_3\right)\left(2\,[F'_1]^{\mu}_{\ph\mu\nu}-[F'_2]^{\mu}_{\ph\mu\nu}-[F'_3]^{\mu}_{\ph\mu\nu}\right)-e_1^2e_3\left(2\,[H'_1]^{\mu}_{\ph\mu\nu}-[H'_2]^{\mu}_{\ph\mu\nu}-[H'_3]^{\mu}_{\ph\mu\nu}\right) \nn\\ 
& +\left(-e_2e_4-e_1^2e_4+e_1e_2e_3\right)\left(2\,[I'_1]^{\mu}_{\ph\mu\nu}-[I'_2]^{\mu}_{\ph\mu\nu}-[I'_3]^{\mu}_{\ph\mu\nu}\right) +\left(e_3e_4-e_1e_3^2\right)\left([J'_1]^{\mu}_{\ph\mu\nu}-[J'_2]^{\mu}_{\ph\mu\nu}\right) = 0 \,.\nn
\end{align}
From $ x_0\,x_2\,x_3^{\ph 3 2} $ we get
\begin{align}
& \left(-2e_2+3e_1^2\right)\left([A'_1]^{\mu}_{\ph\mu\nu}-[A'_2]^{\mu}_{\ph\mu\nu}\right)-e_1e_2\left(2\,[B'_1]^{\mu}_{\ph\mu\nu}-[B'_2]^{\mu}_{\ph\mu\nu}-[B'_3]^{\mu}_{\ph\mu\nu}\right) \nn\\
&+\left(-2e_4+2e_1e_3\right)\left(2\,[C'_1]^{\mu}_{\ph\mu\nu}-[C'_2]^{\mu}_{\ph\mu\nu}-[C'_3]^{\mu}_{\ph\mu\nu}\right)  \nn\\ 
& +\left(-2e_4+2e_1e_3+2e_2^2-2e_1^2e_2\right)\left([D'_1]^{\mu}_{\ph\mu\nu}-[D'_2]^{\mu}_{\ph\mu\nu}\right) \nn\\
&+\left(e_1^2e_3-e_2e_3-e_1e_4\right)\left(2\,[E'_1]^{\mu}_{\ph\mu\nu}-[E'_2]^{\mu}_{\ph\mu\nu}-[E'_3]^{\mu}_{\ph\mu\nu}\right) \nn\\ 
& +\left(-3e_1e_4+e_2e_3\right)\left(2\,[F'_1]^{\mu}_{\ph\mu\nu}-[F'_2]^{\mu}_{\ph\mu\nu}-[F'_3]^{\mu}_{\ph\mu\nu}\right)+\left(2e_2e_4-e_1e_2e_3\right)\left(2\,[H'_1]^{\mu}_{\ph\mu\nu}-[H'_2]^{\mu}_{\ph\mu\nu}-[H'_3]^{\mu}_{\ph\mu\nu}\right) \nn\\ 
& +\left(-2 e_1e_2e_4+e_2^2e_3\right)\left(2\,[I'_1]^{\mu}_{\ph\mu\nu}-[I'_2]^{\mu}_{\ph\mu\nu}-[I'_3]^{\mu}_{\ph\mu\nu}\right) +\left(2e_4^2+e_1e_3e_4-e_2e_3^2\right)\left([J'_1]^{\mu}_{\ph\mu\nu}-[J'_2]^{\mu}_{\ph\mu\nu}\right) = 0 \,.\nn
\end{align}
From $ x_0\,x_3^{\ph 3 3} $ we get
\begin{align}
& \left(e_1^3-e_1e_2-e_3\right)\left([A'_1]^{\mu}_{\ph\mu\nu}-[A'_2]^{\mu}_{\ph\mu\nu}\right)+\left(-e_4+e_1e_3+e_2^2-e_1^2e_2\right)\left(2\,[B'_1]^{\mu}_{\ph\mu\nu}-[B'_2]^{\mu}_{\ph\mu\nu}-[B'_3]^{\mu}_{\ph\mu\nu}\right) \nn\\ 
& +\left(-e_1e_4-e_2e_3+e_1^2e_3\right)\left(2\,[C'_1]^{\mu}_{\ph\mu\nu}-[C'_2]^{\mu}_{\ph\mu\nu}-[C'_3]^{\mu}_{\ph\mu\nu}\right) +\left(-e_1^2e_4+e_3^2\right)\left(2\,[F'_1]^{\mu}_{\ph\mu\nu}-[F'_2]^{\mu}_{\ph\mu\nu}-[F'_3]^{\mu}_{\ph\mu\nu}\right) \nn\\ 
& +\left(e_3e_4+e_1e_2e_4-e_1e_3^2\right)\left(2\,[H'_1]^{\mu}_{\ph\mu\nu}-[H'_2]^{\mu}_{\ph\mu\nu}-[H'_3]^{\mu}_{\ph\mu\nu}\right) \nn\\
&+\left(e_4^2-e_1e_3e_4-e_2^2e_4
+e_2e_3^2\right)\left(2\,[I'_1]^{\mu}_{\ph\mu\nu}-[I'_2]^{\mu}_{\ph\mu\nu}-[I'_3]^{\mu}_{\ph\mu\nu}\right) \nn\\ 
& +\left(e_1e_4^2+e_2e_3e_4-e_3^3\right)\left([J'_1]^{\mu}_{\ph\mu\nu}-[J'_2]^{\mu}_{\ph\mu\nu}\right) = 0 \,.\nn\\
\end{align}
Here we find that only the four following syzygies are independent
\begin{align}\label{syz1}
&[A'_1]^{\mu}_{\ph\mu\nu}-[A'_2]^{\mu}_{\ph\mu\nu}-e_2\left([D'_1]^{\mu}_{\ph\mu\nu}-[D'_2]^{\mu}_{\ph\mu\nu}\right) -e_3\left(2[F'_1]^{\mu}_{\ph\mu\nu}-[F'_2]^{\mu}_{\ph\mu\nu}-[F'_3]^{\mu}_{\ph\mu\nu}\right)\nn\\
&-\left(e_4-e_1e_3\right)\left([G'_1]^{\mu}_{\ph\mu\nu}-[G'_2]^{\mu}_{\ph\mu\nu}\right)
-\left(e_4-e_1e_3\right)\left(2[H'_1]^{\mu}_{\ph\mu\nu}-[H'_2]^{\mu}_{\ph\mu\nu}-[H'_3]^{\mu}_{\ph\mu\nu}\right)\nn\\
&-e_2e_3\left(2[I'_1]^{\mu}_{\ph\mu\nu}-[I'_2]^{\mu}_{\ph\mu\nu}-[I'_3]^{\mu}_{\ph\mu\nu}\right) +e_3^2\left([J'_1]^{\mu}_{\ph\mu\nu}-[J'_2]^{\mu}_{\ph\mu\nu}\right) = 0 \,,\\[10pt]\label{syz2}
& 2[B'_1]^{\mu}_{\ph\mu\nu}-[B'_2]^{\mu}_{\ph\mu\nu}-[B'_3]^{\mu}_{\ph\mu\nu} -e_1\left([D'_1]^{\mu}_{\ph\mu\nu}-[D'_2]^{\mu}_{\ph\mu\nu}\right)+e_3\left([G'_1]^{\mu}_{\ph\mu\nu}
-[G'_2]^{\mu}_{\ph\mu\nu}\right)\nn\\
&-e_4\left(2[I'_1]^{\mu}_{\ph\mu\nu}-[I'_2]^{\mu}_{\ph\mu\nu}
-[I'_3]^{\mu}_{\ph\mu\nu}\right) = 0 \,,\\[10pt]\label{syz3}
& 2[C'_1]^{\mu}_{\ph\mu\nu}-[C'_2]^{\mu}_{\ph\mu\nu}-[C'_3]^{\mu}_{\ph\mu\nu} +\left([D'_1]^{\mu}_{\ph\mu\nu}-[D'_2]^{\mu}_{\ph\mu\nu}\right)+e_1\left(2[F'_1]^{\mu}_{\ph\mu\nu}
-[F'_2]^{\mu}_{\ph\mu\nu}-[F'_3]^{\mu}_{\ph\mu\nu}\right)\nn\\
&-\left(e_1^2-e_2\right)\left([G'_1]^{\mu}_{\ph\mu\nu}-[G'_2]^{\mu}_{\ph\mu\nu}\right)
-e_1^2\left(2[H'_1]^{\mu}_{\ph\mu\nu}-[H'_2]^{\mu}_{\ph\mu\nu}-[H'_3]^{\mu}_{\ph\mu\nu}\right)\nn\\ &+e_1e_2\left(2[I'_1]^{\mu}_{\ph\mu\nu}-[I'_2]^{\mu}_{\ph\mu\nu}-[I'_3]^{\mu}_{\ph\mu\nu}\right)
-\left(e_1e_3+e_4\right)\left([J'_1]^{\mu}_{\ph\mu\nu}-[J'_2]^{\mu}_{\ph\mu\nu}\right) = 0 \,,\\[10pt]\label{syz4}
& 2[E'_1]^{\mu}_{\ph\mu\nu}-[E'_2]^{\mu}_{\ph\mu\nu}-[E'_3]^{\mu}_{\ph\mu\nu} +\left(2[F'_1]^{\mu}_{\ph\mu\nu}-[F'_2]^{\mu}_{\ph\mu\nu}-[F'_3]^{\mu}_{\ph\mu\nu}\right) 
-e_1\left([G'_1]^{\mu}_{\ph\mu\nu}-[G'_2]^{\mu}_{\ph\mu\nu}\right)\nn\\&
-e_1\left(2[H'_1]^{\mu}_{\ph\mu\nu}-[H'_2]^{\mu}_{\ph\mu\nu}-[H'_3]^{\mu}_{\ph\mu\nu}\right)  +e_2\left(2[I'_1]^{\mu}_{\ph\mu\nu}-[I'_2]^{\mu}_{\ph\mu\nu}-[I'_3]^{\mu}_{\ph\mu\nu}\right)-e_3\left([J'_1]^{\mu}_{\ph\mu\nu}-[J'_2]^{\mu}_{\ph\mu\nu}\right) = 0 \,.
\end{align}
By contracting these equations with $\nabla_\mu \nabla^\nu$ we obtained the 4 relations (\ref{SY1})-(\ref{SY2}) of the main text.

\section{Another way to the linearized field equations}
\label{otherway2}
We could as well have derived the field equations directly from the background equation in the form  (\ref{b1_Ssol})
\ba {S_{\mu\nu}=-\frac{\beta_0}{3\beta_1}g_{\mu\nu}+\frac{1}{m^2\beta_1}\left(R_{\mu\nu}-\frac{1}{6}g_{\mu\nu}R\right)}.\ea 
Indeed we can read this equation as expressing the metric $ f_{\mu\nu} $ as function of the metric $g_{\mu \nu}$ and its curvature as 
\be
f_{\mu\nu}=\dfrac{\beta_0^2}{9\beta_1^2}g_{\mu\nu}-\dfrac{2\beta_0}{3m^2\beta_1^2}\left(R_{\mu\nu}-\frac{1}{6}R g_{\mu\nu}\right)+\dfrac{1}{m^4\beta_1^2}\left(R_{\mu\rho}R^{\rho}_{\ph\rho\nu}+\frac{1}{36}R^2g_{\mu\nu}-\frac{1}{3}RR_{\mu\nu}\right) \,,
\ee
and then use the fact that the variation $ \delta f_{\mu\nu}$ of $ f_{\mu\nu} $ vanishes, since $ f_{\mu\nu} $ is not a dynamical variable, to obtain the field equations as an equation for $ \delta G_{\mu\nu}$ taking the form
\be
A_{\mu\nu}^{\ph\mu\ph\nu\rho\sigma}\,\delta G_{\rho\sigma}=T_{\mu\nu} \,,
\ee
with
\be
A_{\mu\nu}^{\ph\mu\ph\nu\rho\sigma} = \dfrac{2}{3}m^2\beta_0\left(\delta^{\rho}_{\mu}\delta^{\sigma}_{\nu}-\dfrac{1}{3}g_{\mu\nu}g^{\rho\sigma}\right)-2R^{\rho}_{(\mu}\delta^{\sigma}_{\nu)}+\dfrac{1}{3}R\,\delta^{\rho}_{\mu}\delta^{\sigma}_{\nu}+\dfrac{2}{3}R_{\mu\nu}g^{\rho\sigma}-\dfrac{1}{9}R\,g_{\mu\nu}g^{\rho\sigma} \,,  
\ee
and 
\begin{align}
T_{\mu\nu} =&\, \dfrac{1}{9}m^4\beta_0^{\ph 0 2}\,h_{\mu\nu}-\dfrac{2}{9}m^2\beta_0 R^{\rho\sigma}h_{\rho\sigma}g_{\mu\nu}+\dfrac{1}{9}m^2\beta_0 R\,h\,g_{\mu\nu}-\dfrac{2}{9}m^2\beta_0 R\,h_{\mu\nu }+\dfrac{2}{3}R^{\rho\sigma}h_{\rho\sigma}R_{\mu\nu} \\
\nonumber
&-\dfrac{1}{3}h\,R\,R_{\mu\nu}-\dfrac{1}{9}R\,R^{\rho\sigma}h_{\rho\sigma}g_{\mu\nu}+\dfrac{1}{18}R^2\,h\,g_{\mu\nu}+RR^{\rho}_{(\mu}h_{\nu)\rho}-\dfrac{5}{36}R^2\,h_{\mu\nu}-R^{\rho}_{\mu}R^{\sigma}_{\nu}\,h_{\rho\sigma}  \,.
\end{align}
We then invert formally the operator $A$ defining $B$ that verifies 
\ba
A_{\mu\nu}^{\ph\mu\ph\nu\rho\sigma}B_{\rho\sigma}^{\ph\rho\ph\sigma\lambda\gamma}=\frac{1}{2}\left(\delta^{\lambda}_{\mu}\delta^{\gamma}_{\nu}+\delta^{\gamma}_{\mu}\delta^{\lambda}_{\nu}\right).
\ea
This yields 
\ba
\delta G_{\mu\nu} = B_{\mu\nu}^{\ph\mu\ph\nu\rho\sigma}T_{\rho\sigma}.
\ea
which should agree with the field equations~\eqref{fieldsummary} obtained previously where we identify the mass matrix as,
\be
m^2\mathcal{M}_{\mu\nu}^{\ph\mu\ph\nu\rho\sigma}h_{\rho\sigma} = - B_{\mu\nu}^{\ph\mu\ph\nu\rho\sigma}T_{\rho\sigma} \,.
\ee
Thus the following identity should hold:
\be\label{inv}
 T_{\mu\nu} = - m^2A_{\mu\nu}^{\ph\mu\ph\nu\rho\sigma}\mathcal{M}_{\rho\sigma}^{\ph\rho\ph\sigma\lambda\gamma}h_{\lambda\gamma}  \,.
\ee
To check it, we first use once again the background equation to write everything in terms of $ S $ and find that $ A $ can be written in the simple form,
\be
A_{\mu\nu}^{\ph\mu\ph\nu\rho\sigma} = m^2\beta_1\left(-2S^{\rho}_{(\mu}\delta^{\sigma}_{\nu)}+\dfrac{2}{3}S_{\mu\nu}g^{\rho\sigma}\right) \,.
\ee
We then use the expression of $ \mathcal{M} $ in the $\beta_1$ model, reduce all powers of $ S $ greater than three and use the equations~\eqref{syz1}-\eqref{syz4} to finally get that the equation~\eqref{inv} is indeed verified, both sides of this equation being,
\begin{align}
T_{\mu\nu} &= -m^2A_{\mu\nu}^{\ph\mu\ph\nu\rho\sigma}\mathcal{M}_{\rho\sigma}^{\ph\rho\ph\sigma\lambda\gamma}
h_{\lambda\gamma} \nn\\
&= m^2\beta_1\left[2m^2\left(\beta_0+\beta_1 e_1\right)S^{\rho}_{(\mu}h_{\nu)\rho}-\dfrac{2}{3}m^2\left(\beta_0+\beta_1 e_1\right)h\,S_{\mu\nu}-m^2\beta_1 S^{\rho}_{\mu}S^{\sigma}_{\nu}h_{\rho\sigma} \right.\nn\\
&\qquad\qquad\left.
+\dfrac{2}{3}m^2\beta_1 S^{\rho\sigma}h_{\rho\sigma}S_{\mu\nu}\right] \,.
\end{align}


\end{document}